\documentstyle[11pt,aaspp4]{article}

\begin{document}

\title{The Temperature and Opacity of Atomic Hydrogen 
in Spiral Galaxies}

\author{Robert Braun}
\affil{Netherlands Foundation for Research in Astronomy, Postbus~2, 
7990~AA Dwingeloo, The Netherlands}

\begin{abstract}
  
  We analyze the resolved neutral hydrogen emission properties of a
  sample of eleven of the nearest spiral galaxies. Between 60 and 90\%
  of the total \ion{H}{1} line flux within the optical disk is due to
  a high brightness network (HBN) of emission features which are
  marginally resolved in their narrow dimension at about 150~pc and
  have a face-on covering factor of about 15\%. Averaged line profiles
  of this component are systematically non-Gaussian with a narrow core
  (less than about 6~km~s$^{-1}$ FWHM) superposed on broad Lorentzian
  wings (30~km~s$^{-1}$ FWHM). An upper limit to the gas temperature
  of 300~K follows directly from the narrow line profiles, while
  simple modeling suggests kinetic temperatures equal to the peak
  emission brightness temperature (80--200~K) in all cases but the
  outer disks of low mass galaxies, where the HBN becomes optically
  thin to the $\lambda$21~cm line. Positive radial gradients in the
  derived kinetic temperature are found in all spiral galaxies.  The
  distributions of brightness temperature with radius in our sample
  form a nested system with galaxies of earlier morphological type
  systematically displaced to lower temperature at all radii. The
  fractional line flux due to the HBN plummets abruptly near the edge
  of the optical disk where a diffuse outer gas disk takes over. We
  identify the HBN with the Cool Neutral Medium.

\end{abstract}


\section{Introduction}

The last three decades have seen substantial progress in our
understanding of the neutral interstellar medium of galaxies, as
summarized for example in the review by Kulkarni \& Heiles
(1988). Even so, there are important elements of the global theory
which are still rather poorly defined. One of the most basic of these
is the intrinsic topology of the neutral gas. Historically, it was
convenient to characterize the neutral ISM in the solar neighborhood
as a distribution of ``standard'' diffuse spherical clouds of atomic
hydrogen with typical radii of perhaps 5~pc (Spitzer 1978). This was
done even though earlier observational studies (Heiles 1967, Verschuur
1974) had demonstrated that the majority of atomic structures were
organized into large sheets or filaments with dimensions of at least
tens of pc. The basic question has remained, however, of whether the
atomic gas topology more closely resembles isolated concentrations in
a lower density substrate (the ``raisin pudding'' model), a moderate
density matrix with low density cavities (the ``Swiss cheese'' model)
or perhaps a self-similar distribution of structures over some range
of spatial scales.  There is also the question of whether this
topology might depend on location within the galactic disk.

By turning to nearby external galaxies it becomes much easier to
achieve a global inventory of the neutral gas topology. An entire
galactic disk, at a common distance and more favorable orientation, can
then be studied simultaneously. The only drawback to this approach is
the difficulty of obtaining sufficient physical resolution and
sensitivity. This drawback has been serious enough that it has
precluded serious study of the neutral atomic ISM of external galaxies
until rather recently. The many published studies of \ion{H}{1}
emission in external galaxies have instead concentrated primarily on
the global gas dynamics and kinematics. And even though 100~pc by
8~km~s$^{-1}$ resolution was available about ten years ago in
\ion{H}{1} studies of the Local Group galaxies M31 and M33 (Brinks \&
Shane 1984, Deul \& van der Hulst 1987) the question of global topology
of the neutral ISM in these systems has not yet been addressed. What
has emerged, from these and more recent studies, is the recognition
that bubble and filamentary emission structures are indeed the norm
rather than the exception.

A related fundamental concern is the physical state of the gas itself.
What are the typical densities and temperatures involved and do these
properties depend in any systematic way on position in a galaxy.
Theoretical treatments of \ion{H}{1} thermodynamics (Field 1965,
Draine 1978, Shull \& Woods 1985, Wolfire et al. 1995) lead to the
clear prediction that two distinct phases of neutral atomic hydrogen
are to be expected. A warm phase, the Warm Neutral Medium (WNM), with
density of order 0.1~cm$^{-3}$ and temperature of about 8000~K should
predominate at low interstellar pressures, while a cool phase, the
Cool Neutral Medium (CNM) with density $>$~1~cm$^{-3}$ and temperature
between about 50 and 250~K should become dominant at high interstellar
pressures. Strong observational evidence for these two distinct
components dates back to some of the earliest comparisons of
\ion{H}{1} emission and absorption along Galactic lines-of-sight
(Clark 1965, Radhakrishnan et al. 1972). However, given the paucity of
background sources with which to map out the distribution of
absorption, it has remained difficult to associate specific spatial
morphologies to the two atomic phases, or to track any possible
variation in their physical parameters with location in the Galaxy.

This problem can also be usefully addressed by turning to nearby
external systems, where a more favorable orientation can facilitate
unambiguous identification of particular components and allow
determination of their location within a galaxy. The scarcity of
suitable background sources for the measurement of \ion{H}{1}
absorption in the direction of nearby galaxy disks makes this a
difficult undertaking. Even so, a comparison of \ion{H}{1} emission and
absorption spectra through the disk of M31 with 35~pc by 5~km~s$^{-1}$
resolution (Braun \& Walterbos 1992) reveals that significant
absorption opacities have a one-to-one correspondence with regions of
high emission brightness. Although this may sound like a trivial
statement, the implications are quite profound. {\it \ion{H}{1}
emission structures with a high brightness temperature are relatively
opaque, so that the emission brightness approaches the gas kinetic
temperature when the structure has been spatially and kinematically
resolved.} Furthermore, the cool phase kinetic temperature in M31 is
observed to increase systematically from about 70~K at radii of
5--10~kpc to about 175~K at radii of 10--20~kpc. The identical positive
radial gradient is observed directly in the peak emission brightness
temperature! This close correspondence of emission brightness and
kinetic temperature implies that the high brightness emission features:
(1) have been resolved (at about 100~pc by 5~km~s$^{-1}$) and (2) have
\ion{H}{1} opacities greater than about unity over the range in radius
where the temperatures agree.

In this paper we will analyze the \ion{H}{1} emission properties,
observed with a resolution of about 100~pc by 6~km~s$^{-1}$, of a
sample of eleven of the nearest spiral galaxies beyond the Local Group.
The basic data have been presented in Braun (1995, hereafter referred
to as B95) so only the most relevant points concerning the observations
and data reduction are summarized below in \S~2. An important point to
stress at the outset is that although this database has high spatial
resolution it also has full sensitivity to the largest structures and
accurately recovers the total integrated emission of each galaxy. High
resolution images of both the peak and the integrated brightness of all
the sample galaxies are presented in B95. Images of the peak brightness
of \ion{H}{1} emission at this physical resolution are particularly
illuminating, in that they reveal the existence of a filamentary
network of high brightness temperature ($>$ 50~K) emission features
within the star-forming disk of each galaxy. These features include,
but are not restricted to, the spiral arms seen in optical,
molecular and dust tracers. 

In the few cases where the presence of an appropriate background
source has allowed a search for \ion{H}{1} absorption, gas kinetic
temperatures are implied which are consistent with the observed
brightness temperature, just as was seen previously in the case of
M31. This high brightness network (or HBN) gives every indication of
representing a phase of the atomic gas which has substantial
\ion{H}{1} opacity. An interesting pattern was noted in B95 for the
peak brightness temperature of these features to increase
systematically with radius. It seems that there may be a systematic
variation with radius of either or both of (1) the gas kinetic
temperature and (2) the typical \ion{H}{1} opacity of the high
brightness \ion{H}{1} in normal spiral galaxy disks.

In \S~3 we will derive the properties of the high brightness
\ion{H}{1} network as a function of radius in our sample
galaxies. Critical topological and physical properties such as the
spatial scale-size, velocity width, fractional \ion{H}{1} line flux,
surface covering factor and peak brightness temperature will be
determined together with their radial dependencies. In addition,
median co-aligned emission spectra will be generated and analyzed
within annular zones. The extremely narrow line-width of the high
brightness emission core allows an upper limit of about 300~K to be
placed on the gas kinetic temperature. There can be no doubt that the
high brightness network represents high opacity concentrations of the
cool atomic phase of neutral hydrogen.

\section{Observations and Data Reduction}

Neutral hydrogen observations of the eleven program galaxies were
obtained with the VLA between March 1989 and November 1990. The B, C
and D configurations (with effective integration times of about 7, 0.5
and 0.4 hours) were utilized to image a region 0.5 degree in diameter
at 6 arcsec resolution for each of the eight Northern galaxies. In
addition, a small hexagonal mosaic in the D configuration (with 0.4
hours effective integration on each of 7 positions) was used to image a
1 degree diameter field at 65 arcsec resolution. Similar resolution and
sampling was obtained for the three southerly galaxies by observing in
the BnA, CnB and DnC configurations. Observing dates, field centers and
other particulars are summarized in Table 1. Namely, the galaxy name in
column (1), the B1950 pointing center in (2), the observation dates for
the three observed configurations in (3), the central velocity and
number of frequency channels in (6) and (7). The assumed inclination,
position angle of receding line-of-nodes and major axis radius at which
the blue optical surface brightness is 25 mag~arcsec$^2$ (from de
Vaucouleurs et al., 1991, hereafter the RC3) is
given in columns (8), (9) and (10). The galaxy type, approximate
distance, total blue magnitude and luminosity are given in columns (11)
to (14). Standard calibration and imaging techniques were used to
produce a series of narrow-band images separated by 5.16~km~s$^{-1}$
over a velocity range of 330~km~s$^{-1}$ (or 660~km~s$^{-1}$ when
necessary) centered on the nominal heliocentric systemic velocity of
each galaxy. Since a uniform frequency taper was applied in the
correlator, the effective velocity resolution was 6.2~km~s$^{-1}$.

Data attributes are summarized in Table~2, both at the full resolution
as well as after smoothing to spatial resolutions of 9, 15, 25 and 65
arcsec. The best fitting elliptical Gaussian beam parameters and the
beam integral at full spatial resolution are listed in columns (2) and
(3). The rms sensitivity in a single frequency channel is given in
terms of flux density and surface brightness in columns (4) and (5).
The surface brightness sensitivity after convolution to lower spatial
resolutions is given in columns (6) through (9) of the table. Total
detected line fluxes within the contiguous region bounded by a peak
brightness of 4 and 0~Kelvin (as seen at 65~arcsec resolution) are
listed in columns (10) and (11). Recent single dish measurements in the
literature (as tabulated in Huchtmeier and Richter, 1989) generally lie
between these two values.

The line-of-sight emission and absorption properties toward the 54
brightest continuum sources in the 11 observed fields were derived and
are presented in Table 3 of B95.

\section{Results}

Examination of the peak \ion{H}{1} brightness images in Figs.~1--10 of
B95 reveals how the distribution of atomic gas in our sample galaxies
becomes decomposed at 100~pc linear resolution into a high brightness
filamentary network (or HBN) of \ion{H}{1} features. The continuity is best
seen at low to moderate inclinations, while edge-on systems like NGC~55
and 4244 appear entirely filled as seen in projection. This 
network corresponds globally to both the grand-design, as well as the
flocculent, spiral arms traced by massive star formation and by dust
lanes in the various galaxies.

\subsection{ Assessing Data Purity }

Before undertaking an analysis of the HBN, we will demonstrate that we
are dealing with a real population of features rather than some set of
random noise peaks. To begin with, let us consider the noise
properties of the data-cubes. In Fig.~1a we present a histogram of the
pixel brightnesses seen in several velocity channels near the edge of
the observed band toward NGC~5457, where no line emission was
apparent. The noise distribution is quite accurately Gaussian down to
at least plus and minus 4$\sigma$ and is fit by an rms fluctuation
level of 1.05~mJy per beam area. A Gaussian function with this
dispersion is overlaid on the histogram as the solid line. Careful
inspection does reveal a small symmetric excess of pixels between
about 3 and 4$\sigma$ with respect to the $\sigma$~=~1.05 Gaussian
distribution. This portion of the distribution is fit best by the
$\sigma$~=~1.07 Gaussian indicated by the dashed line in the
figure. This difference of 2\% in the rms fluctuation level is
indicative of the accuracy to which the rms can be determined and the
degree to which the data conform to Gaussian statistics.

We will now consider the criteria which permit optimum detection of
emission features. The primary beam corrected image of the peak
observed brightness temperature in NGC~5457 is shown in Fig.~2a. It is
clear that the signal-to-noise ratio in our full resolution data is
not overwhelming. The typical brightness sensitivity is 15--20~K at
the field center, while peak brightnesses on the filamentary
ridgelines vary between about 50 and 200~K. We have isolated the high
brightness network from the general field by applying a spatial mask
to the full resolution images of peak brightness. This mask was
defined by the regions within the 5$\sigma$ contour in the 15~arcsec
resolution images of peak brightness. This cut-off level corresponds
to about 25~K at 15~arcsec resolution at the field center and climbs
to 50~K at a radius of about 1800~arcsec (due to the primary beam
correction). The masked version of the NGC~5457 image is shown in
Fig.~2b to illustrate the effect of applying this mask.

The degree of data purity within the masked database is illustrated in
Fig.~2c, where we plot the line-of-sight velocity which corresponds to
each pixel shown in Fig.~2b. It is important to stress that we are not
simply displaying a velocity field which has been generated by
calculating a weighted mean along each spectrum of the data-cube,
but rather the actual velocity which corresponds to the single pixel
with highest brightness at each position. Random noise peaks will have
a random velocity associated with them drawn from the entire range of
observed velocities. The distribution of velocities in Fig.~2c is not
random, but varies systematically in accordance with the actual
kinematics of the galaxy. A small, but visible contribution of random
velocities is indeed visible, particularly near the edges of the
masked distribution. 

We can obtain a quantitative estimate of the degree of noise
contamination by producing a histogram of the velocities near either
end of the major axis of the galaxy where the galactic kinematics
dictate an approximately constant line-of-sight velocity for real
emission features. In Fig.~1b we present the histogram of observed
velocities (shown in Fig.~2c) for a sub-field centered on the
South-West major axis of NGC~5457. The number density of pixels having
an aberrant velocity (about 100 per velocity channel) implies that as
many as 23\% of the (21,000) pixels isolated by the spatial mask in
this sub-field are likely to be noise interlopers. If we apply the
additional constraint that the peak brightness associated with each
velocity must exceed 3$\sigma$ we get the image shown in
Fig.~2d. Inspection shows that this velocity field is substantially
cleaner and examination of the corresponding histogram of the same
major axis sub-field in Fig.~1c reveals that 5.2\% of the pixels
(relative to the area isolated by the spatial mask) are likely to be
noise peaks given the number density of aberrant velocities (about 20
per velocity channel).

This can be compared with the expectation based on perfectly Gaussian
noise. In that case we would expect to have 0.13\% of the values in a
single image exceed 3$\sigma$. Since the peak brightness image has
been derived from a cube containing about 40 {\it independent}
velocity channels, we would expect it to suffer from a 5.4\%
contamination by random noise peaks exceeding the 3$\sigma$
level. This is in excellent agreement with the estimate derived
above. Constraining the selection of pixels to only those for which
the peak brightness exceeds the 4$\sigma$ level leaves less than
0.27\% contamination of the mask area by random noise peaks as
illustrated in Fig.~1d. The expectation based on purely Gaussian noise
is that only 0.13\% of the random noise peaks should exceed this level
in an image of peak brightness based on 40 independent velocity
channels. This discrepancy with respect to the Gaussian value would
arise if we had underestimated the true rms fluctuation level by only
3\%. Both the magnitude and sense of this discrepancy are in agreement
with the actual noise histogram shown in Fig.~1a.

In practical terms then, of the 85,000 pixels isolated by the spatial
mask in NGC~5457, 8895 are found to have peak brightnesses exceeding
4$\sigma$, but 230 of these are likely to be noise peaks. This
practical example illustrates the utility of applying the spatial mask
to the data as a first step in isolating real emission features. If
the entire image of 10$^6$ pixels were being employed, the 8665 real
emission peaks would be contaminated by an additional 2830 noise peaks
exceeding 4$\sigma$. This expectation is confirmed to better than 1\%
accuracy by a direct measurement of the number of peaks exceeding
4$\sigma$ in the entire image. 

For the sake of completeness, we should also point out the potential
pitfalls of employing such a spatial mask based on a high significance
level in a lower resolution image. The lower resolution mask will
encompass the brightest, compact emission features only if they are
spatially associated with more extended regions of emission. Any truly
isolated compact peaks which have only marginal significance will be
lost by this method, since the signal in the smoothed image will
decline more rapidly than the noise level. For our purposes, this is a
risk worth taking since the factor of ten improvement in data purity
is essential, even at the expense of losing some real emission features.

We are now in a position to calculate the actual degree of noise
contamination which effects data satisfying our selection critera of
(1) lying within the spatial mask defined above, and (2) having a peak
brightness in the spectrum exceeding 4$\sigma$. The contamination
level can be quite simply described by the ratio of the expected
number of noise peaks exceeding 4$\sigma$ within the spatial mask
(using the measured contamination rate of 0.27\% over 40 independent
channels) to the number of such features which are actually observed.
In the case of NGC~5457 we derive a contamination level of
2.6\%. Contamination levels were calculated for all of the analyzed
galaxies and are listed in column (14) of Table~2. These vary from 1\%
in the case of NGC~4236 to about 8\% in the case of NGC~7793, but are
typically about 2.5\%.

There can be no doubt that we are dealing with a real population of
features, even though each individual detection has only a modest
signal-to-noise ratio. The properties of the population itself can be
determined with high precision by averaging within 
sub-samples. Measurement noise can be expected to decrease as the
square root of the number of independent samples being combined. With
our knowledge of the noise contamination levels we can also make
accurate predictions of possible systematic effects which these
might give rise to.

\subsection{The Radial Distribution of Peak Brightness}

The mask shown in Fig.~2b is wide enough to include the wings of the
instrumental response and therefore allows measurement of integrated
fluxes from this component. On the other hand, the peak brightness of
filamentary ridge-lines is substantially diluted by a simple average
over the mask. An estimate of the true peak brightness at each radius
was made by forming a histogram of the brightnesses exceeding 4$\sigma$
observed within the masked elliptical annuli and identifying the
brightness at which 80\% of the pixels are included in the
histogram. Changing the histogram
cut-off level to 70\% or even 50\% of the included pixels only resulted in
a modest overall decrease in peak brightness without further
influencing the observed radial trends. Error bars were calculated
from the primary beam corrected rms level appropriate for each annular
zone and the (square root of the) number of contributing image points.

The radial distribution of peak brightness was determined in this way
for each of the low and moderate inclination program galaxies. (No
model-independent radii could be associated with the emission regions
in the highly inclined galaxies NGC~55 and 4244.) These distributions
are shown in panel (a) of Figs.~3--9 together with the average
surface density of integrated HI. Positive radial gradients of the
peak brightness temperature are detected in all galaxies except
NGC~2366. Values vary from typically 80~K at small radii for most
galaxies to more than 200~K at large radii in the case of NGC~5457.

\subsection{Spatial and Velocity Coherence }

Two physically important properties of the HBN are the spatial
scale-size and the velocity dispersion. Visual inspection of Figs.~1--10
of B95 already suggests that when the network is seen relatively
face-on it has a thread-like character which appears to be only
marginally resolved in one dimension with our maximum angular
resolution of about 6~arcsec which corresponds to 100~pc at the
average galaxy distance of 3.5~Mpc. In order to obtain a global
measure of the spatial coherence we have generated images of the ratio
of peak observed brightness after smoothing the data-cubes to 9~arcsec
and 15~arcsec with respect to the peak brightness observed at full
spatial resolution. If our initial datacube is written as I(x,y,v),
and P(x,y) is the image of maximum brightness over all ``v'' at each
spatial pixel, then we are first producing datacubes smoothed to 9 and
15~arcsec, I$_{9''}$(x,y,v) and I$_{15''}$(x,y,v), their corresponding
images of peak brightness, P$_{9''}$(x,y) and P$_{15''}$(x,y), and
finally the ratio images, SC$_{9''}$(x,y)~=~P$_{9''}$(x,y)/P(x,y) and
SC$_{15''}$(x,y)~=~P$_{15''}$(x,y)/P(x,y). 

Only those pixels which were contained in the spatial mask {\it and\ }
had signal to noise greater than 4 (taking account of the primary beam
correction) in the full resolution cube were used in this comparison.
An unresolved linear source would experience a
diminished peak brightness by a factor of about 6.5/9~=~0.72 and
6.5/15~=~0.43 in these two cases. Values of about 0.77 are typically
seen at 9~arcsec resolution and 0.59 at 15~arcsec. The actual spatial
structure is of course more complex than a simple line.  This will
tend to keep the brightness ratio from falling as quickly with spatial
smoothing as it might otherwise. Another consideration is that a faint
diffuse background is present on large angular scales which also
contributes to a systematically higher brightness ratio after
smoothing. In view of these considerations the observed degree of peak
dilution is consistent with the HBN being only marginally resolved in
its narrow dimension with 100~pc spatial resolution.

The radial dependence of diminished peak brightness was determined by
taking the median in elliptical annuli and is illustrated in panel (b)
of Figs.~3--9. It is striking that inside of about R$_{25}$ (the
radius at which the B band surface brightness has declined to
25~mag~arcsec$^{-2}$) the distributions are relatively flat, while at
about this radius a marked systematic decline in the spatial coherence
becomes apparent.  Coherence values seen at large radii are consistent
with linear structures which are {\it unresolved\ } even at the full
100~pc linear resolution.

A similar procedure was employed to obtain a global measure of the
velocity width of the HBN. The full resolution data-cubes were
smoothed in velocity to a resolution of 10.3 and 20.5~km~s$^{-1}$ to
create I$_{10km}$(x,y,v) and I$_{20km}$(x,y,v) and the ratio was
generated of the resulting peak brightnesses to those observed
originally, VC$_{10km}$(x,y)~=~P$_{10km}$(x,y)/P(x,y) and
VC$_{20km}$(x,y)~=~P$_{20km}$(x,y)/P(x,y).

As before, only those pixels within the spatial mask and with signal
to noise greater than 4 in the full resolution database were
used. Isolated, unresolved features in velocity should be diminished
in brightness by a factor of 6.2/10.3~=~0.60 and 6.2/20.5~=~0.30 for
these two cases. In fact, median values of 0.75 were found at
10~km~s$^{-1}$ and 0.54 at 20~km~s$^{-1}$. The radial dependence of
velocity coherence was found to accurately track the spatial coherence
as shown in panel (b) of Figs.~3--9. The same trend noted previously
is also seen in the velocity coherence, namely that near R$_{25}$ a rapid
decline of the coherence with radius is observed. Velocity coherence
values at the largest radii are consistent with isolated spectral
features which are unresolved at the full 6.2~km~s$^{-1}$ velocity
resolution. 

\subsection{Integrated Line Flux and Surface Covering Factor }

The integrated line flux was determined for the HBN in each galaxy
using the spatial mask described and illustrated above (the 5$\sigma$
contour in the 15~arcsec resolution images). The ratio of line flux in
the network to the total line flux (using the 4~K criterion in the
65~arcsec resolution images) is given in column (12) of
Table~2. Between 20 and 85\% of the total line flux is detected within
the high brightness filaments. The ratio of surface area occupied by
these features versus the total surface area of the disk (using the
same masks as for the flux determination above) is given in column
(13) of Table~2. Between about 6 and 50\% of the total disk area is
occupied by the high brightness network. These values represent an
upper limit to the global surface covering factor of the high
brightness network since the spatial isolating mask is wider than the
actual ridge-line.

It is instructive to consider the radial distribution of fractional
flux and surface covering factor of these features, rather than just
considering the global ratios noted above. The fractional line flux of the
HBN relative to the total line flux is plotted as a function of radius
for the program galaxies in panel (c) of Figs.~3--9. It is striking
that within the actively star-forming portion of each galaxy disk
(indicated crudely by R$_{25}$ in each panel) the
fractional line flux in HBN is between about 60 and 90\%. Coinciding
closely with the edge of the star-forming disk is a precipitous
decline in the fractional line flux. This is in contrast to the total
line flux. The accumulated \ion{H}{1} line flux as a function of
radius is also shown in the figure for both the total and HBN
components. While the accumulated HBN flux saturates at R$_{25}$, the
accumulated total flux is still rising almost linearly through about
1.5--2.0 times R$_{25}$.

The face-on surface covering factor is also plotted as a function of
radius in the figures. The fractional surface area of the HBN was
scaled down by the ratio of the average T$_B$ to the peak T$_B$ (as
defined above) in each elliptical annulus to account for the fact
that the mask used to isolate the HBN is wider than the
actual distribution and therefore gives rise to a peak dilution.
Face-on, dilution corrected surface covering factors of the HBN 
are typically about 15\% within R$_{25}$.

\subsection{Median Emission Spectra of the HBN}

Although each spectrum along an individual spatial pixel of our
data-cubes has only a modest signal-to-noise ratio, a glance at Fig.~2
illustrates that there are many independent spectra available for
combined study. We have generated high signal-to-noise spectra for
further analysis by considering annular ellipses in the galaxies
(corresponding to an interval in radius) and forming the median
spectrum of all those pixels satisfying our selection criteria after
aligning them to the velocity of the peak observed brightness. These
spectra are displayed in panel~(d) of Figs.~3--9. The error bars are
calculated from the actual rms fluctuation level at the relevant
radius and the (square root of the) number of contributing points.  In
general, the spectra are characterized by an extremely narrow core
component superposed on broad line wings which often extend over a
velocity interval of 100~km~s$^{-1}$ at significant intensity.

Recall that in \S3.1 we have demonstrated that by applying the spatial
isolating mask and accepting only those pixels with a brightness
exceeding 4$\sigma$ we have insured a high degree of data purity. The
degree of remaining noise contamination is typically 2.5\% as listed
in column (14) of Table~2 for our program galaxies. With this level of
noise contamination, there should be no discernible systematic effect
on the combined spectra. If on the other hand, a large fraction of
noise spectra were included in the combination, there would be an
enhanced central pixel superposed on a background that would average
to zero. The similar character of such a noise bias to that displayed
by the combined spectra was an important reason for the careful
examination of the noise properties of the data already described in
\S3.1.  We are convinced that we do, in fact, understand the noise
properties of the data rather well. Even so, we have attempted to
test for the presence of a significant noise bias by further
constraining the data acceptance criterion to the 4.5, 5 and subsequently
5.5$\sigma$ level. No difference in the resulting line profiles was
found, but only an increasing rms noise level in accordance with the
decreasing number of pixels. 
It is probably worth noting that \ion{H}{1} emission profiles such as
those illustrated in panel (d) of Figs.~3--9 
have never been previously obtained. Only by observing with 100~pc
spatial resolution has it become possible to resolve out both the
overall gradient of galactic rotation (with typical amplitude of about
10~km~s$^{-1}$kpc$^{-1}$) and various systematic discontinuities like
those due to spiral arm shocks. The extremely narrow core of the line
profile, with a dispersion of less than about 2~km~s$^{-1}$, allows an
upper limit of about 300~K to be placed on the kinetic temperature of
the emitting gas. This is a strict upper limit, since some degree of
turbulent broadening may also be present, and any opacity effects will
further broaden the profile. It is probably most appropriate to
characterize the line core as consistent with a thermal linewidth
commensurate with the brightness temperature and only a very modest
degree of additional broadening. The broad pedestal underlying the
line core extends over almost 100~km~s$^{-1}$ and may arise within
either or both of a more turbulent or much warmer component.

We have fit the HBN spectra with a simple physical model consisting of
two types of gaseous components with a sandwich geometry. A central
layer is assumed to have a line opacity which is Gaussian in
velocity,
\begin{equation}
\tau_G(v) = \tau_G(0) e^{-0.5(v/\sigma_G)^2}
\end{equation}
with a thermal velocity dispersion and central opacity given by,
\begin{equation}
\sigma_G = 0.0912 \sqrt{T_C} 
\end{equation}
and
\begin{equation}
\tau_G(0) = {N_{HC} \over 1.83\times 10^{18} {\rm cm}^{-2} T_C \sigma_G \sqrt{2\pi}}
\end{equation}
The outer layers are arbitrarily assumed to have a Lorentzian
dependance of opacity on velocity,
\begin{equation}
\tau_L(v) = {\tau_L(0) \over v^2 + (\Delta_L/2)^2} 
\end{equation}
in terms of the Lorentzian FWHM, $\Delta_L$, and the central opacity
given by,
\begin{eqnarray}
\tau_L(0) = {1\over2} \quad 
{N_{HL}\Delta_L \over 1.83\times 10^{18} {\rm cm}^{-2} T_C 2\pi}
\end{eqnarray}
The observed emission brightness is then given by,
\begin{eqnarray}
T_B(v) = & (T_C- T_Ce^{-\tau_L}) \cr
\ &  + (T_C- T_Ce^{-\tau_G})e^{-\tau_L} \cr
\ &  + (T_C- T_Ce^{-\tau_L})e^{-\tau_G}e^{-\tau_L} 
\end{eqnarray}
in terms of the four variables, T$_C$, $\Delta_L$, N$_{HC}$ and
N$_{HL}$. We have not indicated the explicit velocity dependence of
$\tau_G$ and $\tau_L$ in the equation for T$_B$ for the sake of
compactness. Also note that the total column density of the Lorentzian
component has been divided equally in two parts, one behind and one in
front of the thermal component. All components are assumed to have the
same kinetic temperature, T$_C$. The resulting model spectra of
T$_B(v)$ were smoothed with a 6.2~km~s$^{-1}$ FWHM function to
simulate the observational data. Least squares fits for these model
parameters are listed in Table~3 for the annular zones in each galaxy.
For each model parameter we also list the deviations which lead to a
doubling of $\chi^2$ when the other parameters are held fixed.  The
smoothed solutions are overlaid on the spectra as the solid lines in
panel (d) of Figs.~3--9.

Although we have attempted to fit the data with physically plausible
components, our decomposition is obviously non-unique. For comparison,
we have also carried out numerous other types of decomposition;
ranging from simple two component Gaussian fits to variants of the
Gaussian plus Lorentzian fit in which the Gaussian dispersion was held
fixed at specific values (to simulate a significant turbulent
component) rather than varying in accordance with the kinetic temperature.
Those solutions were discarded since besides not providing
any physical insight, they were also found to have systematic
residuals in the line shoulders and a higher rms deviation. 

Three of our four free parameters were determined quite robustly by
the least squares fitting procedure; namely the kinetic temperature,
T$_C$, together with the Lorentzian linewidth and column density,
$\Delta_L$ and N$_{HL}$. This is because the kinetic temperature is
well-constrained by the core height and width, while the Lorentzian
parameters are well-constrained by the optically thin line wings.
Kinetic temperatures between about 85 and 230~K were derived with a
typical error of about 10\%. It is important to bear in mind that even
the solutions for the lowest kinetic temperatures in Table~3 such as
T$_C$~=~85~K, for which $\sigma_G$~=~0.92~km~s$^{-1}$, are accompanied
by a sufficient peak opacity that the resulting line profile has a
FWHM~$\sim$~5~km~s$^{-1}$, which is comparable to the velocity
resolution of our spectra.  Lorentzian linewidths were found to vary
between about 13 and 42~km~s$^{-1}$. The associated column densities,
N$_{HL}$, were sufficient to account for basically all of the {\it
  apparent\ } column density, N$_{HA}$, which follows from the
assumption of negligible \ion{H}{1} opacity in each spectrum.  The
actual column density associated with the core component, N$_{HC}$,
was much more poorly constrained in those frequent cases when the
implied peak line opacity ($\tau_{max}~=~\tau_G(0)$ listed in Table~3)
was high. As the peak line opacity increases, its impact on the line
shape becomes increasingly subtle due to the exponential dependance in
eqn.~6.  Central opacities in excess of about 5, should be regarded as
formal fit solutions, but should not be taken literally. The quoted
errors on both N$_{HC}$ and $\tau_{max}$ reflect this uncertainty.

Cool gas in the line core sometimes accounts for only 20\% of the
total \ion{H}{1} column as found in the case of NGC~2366 or as much as
a factor of 10 more than in the line wings as found for NGC~5457.
Although the individiual values of N$_{HC}$ are very uncertain for
large $\tau_{max}$, the general trends do seem significant.
High line opacities are derived at all radii in the cases of NGC~2403,
3031 and 5457 and in the inner disks of NGC~247, 4236 and 7793.  Some
galaxies seem to be characterized by rather opaque filaments which can
harbor a substantial quantity of ``hidden'' atomic gas.  More accurate
modeling of the intrinsic lineshape would be possible with higher
sensitivity and, in particular, higher velocity resolution data.

The derived variation of T$_C$ with radius is in good agreement with
that of the peak brightness temperature data as shown in panel (a) of
Figs.~3--9, whenever the line core opacity becomes substantial.
Positive radial gradients of T$_C$ are derived in all cases but that
of NGC~2366. Good numerical agreement of the derived T$_C$ with the
observed peak T$_B$ is found at all radii in the cases of NGC~2403,
3031 and 5457 and in the inner disks of NGC~247, 4236 and 7793.
Systematically higher values of T$_C$ relative to the peak T$_B$ are
derived for NGC~2366 and beyond the inner disk of NGC~247, 4236 and
possibly 7793 in conjunction with a lower derived line-core opacity.
The systematic nature of this departure, increasing with radius in
only the lowest mass galaxies of our sample, is very suggestive of a
systematic decrease in the HBN opacity with radius.

\subsection{Median Emission Spectra of the Diffuse Disk}

We attempted to carry out a similar procedure to produce high
signal-to-noise spectra of the faint diffuse emission in inter-arm
regions and beyond the star forming disk of our program galaxies. To
this end, we employed the 65~arcsec resolution data-cubes to extract
the median co-aligned spectra within elliptical annuli after
application of our data selection criteria. In this case, a spatial
mask was specified interactively by defining a contour which enclosed
the contiguous region over which a smoothly varying radial velocity
was observed associated with the pixel having peak brightness along
each spectrum. Subsequently, only those pixels for which the peak
brightness exceeded 4$\sigma$ but was less than 7.5~K were included in
the combination. Noise contamination rates were calculated as before
and were in all cases less than 1.2\%. The upper limit to the accepted
brightness of 7.5~K was chosen after inspection of the peak brightness
images to eliminate as much as possible of the ``bleeding'' of high
brightness features (at this low spatial resolution) into inter-arm
and outer disk zones. Only in a few cases did it prove possible to
isolate such low brightness regions in the inner disks. The best
example of the resulting spectra is given in Fig.~10 for the case of
NGC~5457.

A major difficulty with the interpretation of spectra obtained at this
relatively low resolution (65~arcsec corresponding to about 1~kpc) is
the significant contribution of galactic kinematics to the shape of
the line profile. We have estimated the expected degree of line
broadening within the beam by forming images of the local velocity
gradient, $\nabla(x,y)$, defined by,

\begin{eqnarray}
\nabla(x,y) = &\biggl[\biggl({\mid V(x,y)-V(x+\Delta,y)\mid + \mid
V(x,y)-V(x-\Delta,y)\mid \over 2} \biggr)^2 \cr
\ &+\biggl({\mid V(x,y)-V(x,y+\Delta)\mid + \mid
V(x,y)-V(x,y-\Delta)\mid \over 2} \biggr)^2 \biggr]^{1/2} 
\end{eqnarray}

in terms of the velocity, V(x,y) and a spatial offset
$\Delta$. $\nabla(x,y)$ was generated for $\Delta$~=~75~arcsec and
it's median value was calculated for the pixels contributing to each
spectrum shown in Fig.~10. This gradient, scaled to the beam FWHM by
multiplication with a factor 65/75, is indicated by a horizontal error
bar plotted at the half intensity level inside each spectrum. For a
uniform spatial distribution of gas with a negligible intrinsic
linewidth we would expect an observed FWHM comparable to that
indicated by the bar. It is clear from the figure that the observed
linewidth is often dominated by the effects of beam smearing. This
effect is strongly inclination dependent and only in the case of
NGC~5457 did it seem worthwhile to proceed with further analysis of
the low resolution line profiles.

We fit these profiles with the sum of two optically thin Gaussian
components. The peak brightnesses and velocity dispersions are listed
in Table~4. Errors are quoted which correspond to a doubling of
$\chi^2$ from it's minimum value. About 80\% of the line brightness
(4--5~K) was associated with a broad component with a dispersion
varying between about 9 and 18~km~s$^{-1}$. The remaining 20\% of the
line brightness was attributed to a narrower component with a
dispersion between about 3 and 4~km~s$^{-1}$.  Since beam-smearing
accounts for such a large fraction of the observed linewidth (as
indicated by the horizontal bar in each spectrum) it is very difficult
to attach any physical interpretation to these line parameters.

\subsection{Velocity Field Discontinuity Beyond the HBN }

While a detailed analysis of the atomic gas kinematics is deferred to a
subsequent paper, at least one specific aspect deserves comment at this
time. Comparison of panels (a) and (c) of Figs.~1--10 in B95 reveals
that the edge of the HBN distribution in each galaxy seems to
be accompanied by a systematic discontinuity in the projected velocity
field in the form of a kink in the iso-velocity contours. The typical
magnitude of the discontinuity is between 5 and 15~km~s$^{-1}$, while
its sense appears to be towards higher apparent rotation velocity in
most cases. Only in the case of NGC~2366 does there appear to be a
strong discontinuity in the sense of reduced apparent rotation velocity.
Although most of these discontinuities are sufficiently subtle that they
would not generally be classified as warps (eg. Briggs 1990) they are
striking in their ubiquity. Inspection of high quality velocity fields
and integrated \ion{H}{1} emission images in the literature (eg.
Begeman 1987) suggests that a subtle kinematic discontinuity is a very
general phenomenon associated with the faint outer plateau of atomic
gas beyond the star-forming disk. Although this kinematic signature is
similar to the kink in iso-velocity contours associated with spiral
arms at smaller radii, it is circular (and not spiral) in shape and
occurs at the outer edge of the star-forming disk.

A more complex, but possibly related
pattern is seen in the case of NGC~3031. Here the kinematics of the
diffuse and the high brightness gas to the East of the inner disk are
radically different. In this case an explanation must be sought in
kinematic modelling of the entire tidally distorted gaseous envelope and
condensed disks of the M81/M82 system. Less extreme forms of non-planar
geometries and possibly non-axisymmetric dynamics should suffice to
explain the more subtle kinematic distortions seen in the diffuse gas
of other systems.

\subsection{Absorption Properties}

As discussed in B95, the small number of randomly distributed
background sources brighter than about 5~mJy/beam have only a very low
probability of occuring along a direction which intercepts the HBN of
\ion{H}{1} emission. A handful of tentative detections are presented
there while the detected mean spin temperatures and the more stringent
lower limits are also indicated at the relevant radii in the profiles
of panel (a) of Figs.~3--9. What little data there are supports the
general conclusion that only regions of high brightness have a
relatively high opacity. Furthermore, the gas kinetic temperature
within the inner disk is consistent with the brightness temperature
seen directly in emission, implying \ion{H}{1} opacities of unity or
greater.

\section{Discussion}

Neutral hydrogen emission from the disks of spiral galaxies appears to
be morphologically segregated into two distinct components. These are
(a) a high brightness filamentary network (HBN) which is
marginally resolved in the narrow dimension at about 150~pc and has
velocity FWHM less than about 6~km~s$^{-1}$ and (b) a diffuse
inter-arm and outer disk component. Each component accounts for about
one half of the total line flux, but it is the HBN which is
coextensive with the star-forming disk; where it accounts for 60 to
90\% of the \ion{H}{1} line flux within R$_{25}$, from a region with
face-on surface covering factor of about 15\%. The narrow linewidth of
the HBN profiles allows a strict upper limit of 300~K to be placed on
the gas kinetic temperature. This cool temperature allows the
unambiguous identification of the HBN with the Cool Neutral Medium
(CNM) predicted in theoretical treatments of \ion{H}{1} thermodynamics
in a galactic environment.

Positive radial gradients of the HBN peak brightness temperature are
detected in all of the high and intermediate mass spiral galaxies in
our sample. Only the low mass galaxy, NGC~2366, has a relatively flat
distribution of HBN peak brightness with radius. It is interesting to
compare the distributions of peak brightness observed in the various
galaxies of our sample. They have been plotted together in Fig.~11,
both in terms of the physical radius and in units of the blue stellar
disk scalelength. Different symbols have been used to designate the
different morphological types following the RC3 classifications. It is
striking how the different distributions form an almost perfectly
nested set with the earlier morphological types offset to lower
temperatures at any given radius. The nesting appears to be somewhat
better defined in the plot with physical units for radius rather than
stellar scalelengths. The progression in morphological
type is in this case also a progression in stellar
surface density and almost certainly of the mass surface density. 

Median emission spectra of the HBN display a narrow line core, but
also high velocity wings, extending to at least plus and minus
50~km~s$^{-1}$, which are well fit with a Lorentzian distribution.
High velocity outflows seem to be intimately associated with the
regions of highest gas opacity (as traced by high emission
brightness), presumably due to embedded massive star formation. The
detailed line-shape is found to vary in a systematic way with radius
in each galaxy.  Least squares fits to the line profiles in a simple
radiative transfer model suggest a positive radial gradient of the
\ion{H}{1} kinetic temperature in all cases (excepting NGC~2366). The
\ion{H}{1} line core is fit well with a partially opaque isothermal
gas distribution.  Good agreement is found between the derived kinetic
temperature and the measured peak brightness temperature in all cases
but that of NGC~2366, together with a systematic departure in the
outer disk of NGC~247, 4236 and possibly 7793. In the case of low mass
spiral galaxies, the line-profile fitting suggests that the HBN
becomes increasingly optically thin to $\lambda$21~cm emission at
large radii. 
 
Our modeling of the emission line profiles indicates that the nested
distributions of brightness temperature shown in Fig.~11 correspond to
nested distributions of the \ion{H}{1} kinetic temperature. The
physical conditions responsible for determining the kinetic
temperature (the gas phase metallicity, dust content, radiation field
and interstellar pressure) yield a similar result in the outer disks
of massive galaxies and in the inner disks of low mass
systems. Further modelling should allow at least a crude determination
of the radial variation of these physical properties. Simple considerations
(e.g. Walterbos \& Braun 1996) suggest that the expected radial
variations in the gas phase metallicity, dust content and radiation
field have only a marginal influence on the equilibrium kinetic
temperature of the \ion{H}{1}. The strongest variations in temperature
seem to follow from a radial decline in the mid-plane hydrostatic pressure.

The HBN spectra can be contrasted with the \ion{H}{1} emission spectra
obtained for the solar neighbourhood by Kulkarni and Fich (1985). In
that case, average spectra were formed for both the northern and
southern galactic pole regions (NGP and SGP) using latitudes greater
than 70$^\circ$ from the Bell Laboratories survey data. Although the
NGP data are confused by a strongly infalling component, the SGP data
are quite symmetric about the peak intensity. The peak brightness of
the SGP spectrum is about 4~K and the profile has a FWHM of about
22~km~s$^{-1}$. This linewidth would correspond to a gas kinetic
temperature of about 8500~K if turbulent line broadening were
negligible.  Kulkarni and Fich interpret these spectra as arising in
two or three physical components, of various temperatures and
dispersions. An alternate interpretation might be that we are simply
detecting almost pure Warm Neutral Medium (WNM) when we observe out of
plane \ion{H}{1} emission in the solar neighbourhood. 

Taken together, these two results are consistent with the
interpretation that spiral arms are dominated by high brightness, high
opacity \ion{H}{1} emission from the CNM (with kinetic temperatures in
the range from 80 to 200~K), while inter-arm and presumably outer-disk
regions are characterized by the low brightness emission of the WNM
(with a kinetic temperature of about 8000~K). While this would not be
a very surprising circumstance, it is gratifying that the first
measurements that actually probe the global distribution of physical
conditions in the atomic gas of external galaxies do support this
view.

\acknowledgements

The generous allocation of VLA observing time by the NRAO to carry out
this project is gratefully acknowledged. Special thanks are extended
to John Black, Jayaram Chengalur, Jacqueline van Gorkom, Rob Kennicutt
Ren\'e Walterbos and the referee for reading and commenting on the
manuscript. The National Radio Astronomy Observatory is operated by
Associated Universities, Inc., under cooperative agreement with the
National Science Foundation.

\clearpage
\begin{deluxetable}{lccccccccccccc}
\tablecolumns{14}
\setlength{\tabcolsep}{1pt}
\scriptsize
\tablecaption{Log of Observations \label{tbl-1}}
\tablewidth{0pt}
\tablehead{
\colhead{Galaxy}      &\colhead{R.A.(1950), Dec.(1950)} &
\colhead{B or BnA}    &\colhead{C or CnB} &
\colhead{D or DnC}    &\colhead{V$_{Cen}$} &
\colhead{N$_{Ch}$}    &\colhead{Inc.} &
\colhead{PA}          &\colhead{R$_{25}$} &
\colhead{Type}        &\colhead{Dist.} &
\colhead{B$_T$}       &\colhead{L$_B$} \nl
\colhead{\ }          &\colhead{\ } &
\colhead{\ }          &\colhead{\ } &
\colhead{\ }          &\colhead{(km/s)} &
\colhead{\ }          &\colhead{($^\circ$)} &
\colhead{($^\circ$)}  &\colhead{(")} &
\colhead{\ }          &\colhead{(Mpc)} &
\colhead{(mag)}       &\colhead{(10$^9$L$_\odot$)} \nl
\colhead{(1)}          &\colhead{(2)} &
\colhead{(3)}          &\colhead{(4)} &
\colhead{(5)}          &\colhead{(6)} &
\colhead{(7)}          &\colhead{(8)} &
\colhead{(9)}          &\colhead{(10)} &
\colhead{(11)}          &\colhead{(12)} &
\colhead{(13)}          &\colhead{(14)} 
} 
\startdata
N55 &00 12 24.00 $-$39 28 00.0 &N/A &20-05-89 &20-10-89 &+140 &64 &80
&110 &972 &Sc &2.0 &8.22 &3.2 \nl
N247 &00 44 40.00 $-$21 02 24.0
&04/05-03-89 &20-05-89 &20-10-89 &+160 &64 &75 &170 &600 &Sc &2.5 &9.51
&1.5 \nl
N2366 &07 23 37.00 +69 15 05.0 &19/28-03-89 &29-11-90
&01-12-89 &+100 &64 &58 &39 &228 &SBm &3.3 &11.46 &0.44 \nl
N2403 &07
32 01.20 +65 42 57.0 &19-03-89 &29-11-90 &01-12-89 &+130 &64 &62 &125
&534 &Sc &3.3 &8.89 &4.7 \nl
N3031 &09 51 27.60 +69 18 13.0 &16-03-89
&29-11-90 &01-12-89 &$-$45 &128 &60 &330 &774 &Sb &3.3 &7.86 &12. \nl
N4236 &12 14 21.80 +69 44 36.0 &28-03-89 &29-11-90 &01-12-89 &0 &64 &73
&161 &558 &SBd &3.3 &10.06 &1.6 \nl
N4244 &12 14 59.90 +38 05 06.0
&23-03/02-05-89 &29-11-90 &30-11-89 &+245 &64 &80: &233 &486 &Scd &4
&10.60 &1.4 \nl
N4736 &12 48 32.40 +41 23 28.0 &30-03-89 &29-11-90
&30-11-89 &+305 &64 &33 &305 &330 &Sab &3.8 &8.92 &6.1 \nl
N4826 &12 54
16.90 +21 57 18.0 &22-03-89 &29-11-90 &30-11-89 &+415 &128 &66 &300
&282 &Sab &3.8 &9.37 &4.0 \nl
N5457 &14 01 26.60 +54 35 25.0 &21-03-89
&29-11-90 &01-12-89 &+230 &64 &27 &40 &810 &Sc &6.5 &8.18 &35. \nl
N7793 &23 55 16.00 $-$32 52 06.0 &04/05-03-89 &20-05-89 &20-10-89 &+230
&64 &48 &289 &276 &Sd &3.4 &9.65 &2.5 \nl
\enddata
\end{deluxetable}

\begin{deluxetable}{lcccccccccccccc}
\tablecolumns{14}
\setlength{\tabcolsep}{1pt}
\scriptsize
\tablecaption{Data Attributes and Results \label{tbl-2}}
\tablewidth{0pt}
\tablehead{
\colhead{Galaxy} &\colhead{a b p} &\colhead{$\Omega_B$} &
\colhead{$\Delta$S$_F$}    &\colhead{$\Delta$T$_F$} &
\colhead{$\Delta$T$_9$}    &\colhead{$\Delta$T$_{15}$} &
\colhead{$\Delta$T$_{25}$} &\colhead{$\Delta$T$_{65}$} &
\colhead{$\int F_4 dV$}    &\colhead{$\int F_0 dV$} &
\colhead{I$_{HBN}$/I$_T$}   &\colhead{$\Sigma_{HBN}$/$\Sigma_T$} &
\colhead{Noise}
\nl
\colhead{\ }               &\colhead{(") (") ($^\circ$)} &
\colhead{(a.s.$^2$)}       &\colhead{(mJy/bm)} &
\colhead{(K)}              &\colhead{(K)} &
\colhead{(K)}              &\colhead{(K)} &
\colhead{(K)}              &
\multicolumn{2}{c}{(Jy-km~s$^{-1}$)} &\colhead{\ } &
\colhead{\ }               &\colhead{(\%) }
\nl
\colhead{(1)}          &\colhead{(2)} &
\colhead{(3)}          &\colhead{(4)} &
\colhead{(5)}          &\colhead{(6)} &
\colhead{(7)}          &\colhead{(8)} &
\colhead{(9)}          &\colhead{(10)} &
\colhead{(11)}          &\colhead{(12)} &
\colhead{(13)}          &\colhead{(14)}         
} 
\startdata

N55 &16.0 12.2 $-$15 &220 &3.5 &10.9 &... &... &3.98 &0.93 &1525 &1525 
&0.72 &0.29 &... \nl
N247 &6.58 6.01 $-$10 &61.6 &1.9 &21.1 &12.7 &6.47 &3.78 &0.76 &765 &860 
&0.82 &0.50 &2.4 \nl
N2366 &5.67 5.66 $-$40 &33.8 &1.3 &26.4 &11.2 &4.58 &1.94 &0.73 &235 &250 
&0.69 &0.34 &2.2 \nl
N2403 &6.25 6.12 +66 &48.0 &1.2 &17.2 &9.73 &4.85 &2.52 &0.68 &1320 &1440 
&0.58 &0.26 &2.0 \nl
N3031 &6.13 5.84 $-$1 &45.6 &1.1 &16.6 &8.98 &4.31 &2.52 &0.70 &1455 &1865 
&0.20 &0.074 &4.3 \nl
N4236 &5.93 5.79 $-$38 &49.2 &1.2 &16.8 &8.98 &4.85 &2.52 &0.63 &550 &610 
&0.74 &0.43 &1.0 \nl
N4244 &6.84 5.95 $-$82 &55.6 &1.4 &17.4 &10.5 &5.12 &2.62 &0.57 &410 &435 
&0.85 &0.41 &... \nl
N4736 &6.55 5.87 +85 &48.8 &1.3 &18.3 &9.73 &4.58 &2.23 &0.58 &43 &70 &0.26 
&0.057 &...  \nl
N4826 &6.49 6.10 $-$67 &46.4 &1.3 &19.2 &9.73 &4.58 &2.23 &0.58 &... &47 &... 
&... &... \nl
N5457 &6.04 5.86 +69 &44.8 &1.1 &16.8 &8.23 &4.31 &2.43 &0.62 &1495 &1880 
&0.34 &0.13 &2.6 \nl
N7793 &6.98 6.06 $-$5 &53.2 &2.0 &26.0 &17.2 &8.09 &3.69 &0.82 &210 &270 
&0.41 &0.14 &7.8 \nl
\enddata
\end{deluxetable}

\clearpage
\begin{deluxetable}{lccccccccccccc}
\tablecolumns{8}
\setlength{\tabcolsep}{1pt}
\scriptsize
\tablecaption{Fits to HBN Spectra \label{tbl-3}}
\tablewidth{0pt}
\tablehead{
\colhead{Galaxy}      &\colhead{Radius} &
\colhead{T$_C$}    &\colhead{$\Delta_L$} &
\colhead{N$_{HC}$}    &\colhead{N$_{HL}$} &
\colhead{$\tau_{max}$}& \colhead{N$_{HA}$} \nl
\colhead{\ }          &\colhead{(")} &
\colhead{(K)}          &\colhead{(km/s)} &
\multicolumn{2}{c}{(10$^{21}$cm$^{-2}$)} &\colhead{\ } &
\colhead{(10$^{21}$cm$^{-2}$)} \nl
\colhead{(1)}          &\colhead{(2)} &
\colhead{(3)}          &\colhead{(4)} &
\colhead{(5)}          &\colhead{(6)} &
\colhead{(7)}          &\colhead{(8)}
} 
\startdata
N247 & 75& $120^{+16}_{-15}$& $23^{+7}_{-7}$& $2.5^{+4.0}_{-1.3}$&
$4.0^{+0.5}_{-0.8}$& 4.5$^{+7.1}_{-2.1}$& 3.55  \nl
\    &225& $109^{+6}_{-6}$& $20^{+3}_{-3}$& $6.3^{+6.6}_{-3.7}$&
$5.0^{+0.4}_{-0.5}$& 13$^{+14}_{-7.5}$& 4.03  \nl
\    &375& $115^{+7}_{-7}$& $15^{+3}_{-2}$& $4.0^{+5.4}_{-2.2}$&
$5.0^{+0.4}_{-0.4}$& 7.7$^{+10}_{-4.3}$& 3.93  \nl
\    &525& $156^{+10}_{-10}$& $20^{+3}_{-2}$& $1.6^{+0.3}_{-0.3}$&
$4.0^{+0.3}_{-0.2}$& 1.9$^{+0.4}_{-0.3}$& 3.71  \nl
\    &675& $174^{+12}_{-12}$& $26^{+5}_{-4}$& $2.0^{+0.6}_{-0.3}$&
$4.0^{+0.5}_{-0.3}$& 2.1$^{+0.7}_{-0.3}$& 3.92  \nl
N2366& 50& $180^{+27}_{-27}$& $17^{+4}_{-2}$& $1.0^{+0.5}_{-0.4}$&
$6.3^{+0.7}_{-0.3}$& 1.0$^{+0.5}_{-0.4}$& 5.12  \nl
\    &150& $232^{+40}_{-40}$& $20^{+2}_{-2}$& $1.0^{+0.2}_{-0.2}$&
$5.0^{+0.4}_{-0.4}$& 0.7$^{+0.2}_{-0.2}$& 4.56  \nl
\    &250& $156^{+25}_{-20}$& $20^{+7}_{-6}$& $2.5^{+2.5}_{-0.9}$&
$4.0^{+0.6}_{-0.8}$& 3.1$^{+3.0}_{-1.1}$& 3.84  \nl
N2403&115& $87^{+6}_{-4}$& $15^{+3}_{-2}$& $5.8^{+10}_{-4}$&
$3.9^{+0.3}_{-0.3}$& 17$^{+39}_{-12}$& 3.04  \nl
\    &340& $90^{+5}_{-5}$& $13^{+3}_{-1}$& $4.8^{+8}_{-3}$&
$3.9^{+0.3}_{-0.3}$& 13$^{+25}_{-9}$& 2.99  \nl
\    &565& $94^{+4}_{-4}$& $20^{+5}_{-3}$& $18^{+9}_{-7}$&
$2.7^{+0.2}_{-0.3}$& 50$^{+23}_{-18}$& 2.62   \nl
\    &790& $94^{+5}_{-5}$& $35^{+13}_{-13}$& $28^{+26}_{-10}$&
$2.2^{+0.4}_{-0.4}$& 72$^{+69}_{-26}$& 2.41   \nl
\   &1015& $108^{+11}_{-12}$& $35^{+43}_{-16}$& $11^{+11}_{-7}$&
$1.8^{+0.7}_{-0.7}$&  22$^{+23}_{-15}$& 2.22 \nl
N3031&275& $87^{+9}_{-9}$& $13^{+5}_{-4}$& $4.0^{+10}_{-3}$&
$3.2^{+0.2}_{-0.6}$& 12$^{+42}_{-8}$& 2.58  \nl
\    &385& $90^{+6}_{-5}$& $13^{+4}_{-3}$& $10^{+20}_{-7}$&
$2.5^{+0.6}_{-0.2}$& 28$^{+60}_{-18}$& 2.31   \nl
\    &495& $96^{+5}_{-5}$& $15^{+3}_{-2}$& $2.5^{+2.4}_{-1.0}$&
$3.2^{+0.2}_{-0.2}$& 6.3$^{+6.0}_{-2.5}$& 2.72  \nl
\    &605& $101^{+6}_{-6}$& $20^{+4}_{-4}$& $5.0^{+5.0}_{-2.8}$&
$3.2^{+0.3}_{-0.3}$& 12$^{+12}_{-7}$& 2.90  \nl
\    &715& $104^{+7}_{-7}$& $17^{+6}_{-5}$& $5.0^{+6.5}_{-2.8}$&
$2.5^{+0.3}_{-0.3}$& 11$^{+15}_{-6}$& 2.52  \nl
N4236& 75& $86^{+9}_{-9}$& $13^{+5}_{-3}$& $3.2^{+32}_{-2.6}$&
$4.0^{+0.6}_{-0.6}$& 9$^{+95}_{-8}$& 2.95  \nl
\    &225& $88^{+5}_{-5}$& $13^{+2}_{-2}$& $3.2^{+10}_{-1.9}$&
$4.0^{+0.3}_{-0.3}$& 9$^{+27}_{-5}$& 2.98  \nl
\    &375& $102^{+8}_{-8}$& $15^{+2}_{-2}$& $1.0^{+0.9}_{-0.4}$&
$4.0^{+0.3}_{-0.3}$& 2.3$^{+2.1}_{-0.8}$& 3.12  \nl
\    &525& $123^{+15}_{-9}$& $13^{+2}_{-2}$& $0.8^{+0.4}_{-0.2}$&
$3.2^{+0.3}_{-0.2}$& 1.4$^{+0.6}_{-0.3}$& 2.73  \nl
\    &675& $162^{+73}_{-26}$& $13^{+3}_{-3}$& $0.6^{+0.3}_{-0.2}$&
$2.5^{+0.5}_{-0.2}$& 0.7$^{+0.4}_{-0.2}$& 2.42  \nl
N5457&100& $98^{+15}_{-10}$& $17^{+7}_{-5}$& $2.0^{+4.0}_{-0.9}$&
$2.5^{+0.4}_{-0.5}$& 4.9$^{+12}_{-2.3}$& 2.38  \nl
\    &300& $84^{+8}_{-6}$& $17^{+8}_{-7}$& $32^{+180}_{-26}$&
$3.2^{+0.6}_{-0.6}$& 98$^{+570}_{-78}$& 2.75   \nl
\    &500& $110^{+10}_{-10}$& $15^{+4}_{-2}$& $2.0^{+2.3}_{-0.9}$&
$4.0^{+0.3}_{-0.5}$& 4.2$^{+4.7}_{-2.0}$& 3.28  \nl
\    &700& $127^{+6}_{-6}$& $17^{+3}_{-3}$& $4.0^{+2.1}_{-1.2}$&
$3.2^{+0.3}_{-0.2}$& 6.7$^{+3.5}_{-2.0}$& 3.13  \nl
\    &900& $145^{+10}_{-10}$& $20^{+7}_{-6}$& $6.3^{+6.0}_{-2.7}$&
$4.0^{+0.5}_{-0.2}$& 8.7$^{+8.0}_{-3.7}$& 3.93  \nl
N7793& 50& $129^{+25}_{-25}$& $42^{+65}_{-27}$& $10^{+58}_{-7.5}$&
$3.5^{+1.7}_{-1.6}$& 16$^{+94}_{-12}$& 3.52   \nl
\    &150& $129^{+10}_{-10}$& $17^{+9}_{-5}$& $7.4^{+8.5}_{-4.4}$&
$3.5^{+0.7}_{-0.6}$& 12$^{+14}_{-7}$& 3.44  \nl
\    &250& $150^{+30}_{-30}$& $25^{+21}_{-12}$& $3.5^{+5.4}_{-2.0}$&
$3.0^{+1.4}_{-0.9}$& 4.6$^{+7.1}_{-2.6}$& 3.30  \nl
\enddata
\end{deluxetable}
\clearpage
\begin{deluxetable}{lccccccccccccc}
\tablecolumns{6}
\setlength{\tabcolsep}{1pt}
\scriptsize
\tablecaption{Fits to Diffuse Disk Spectra \label{tbl-4}}
\tablewidth{0pt}
\tablehead{
\colhead{Galaxy}      &\colhead{Radius} &
\colhead{T$_{B1}$}    &\colhead{$\sigma_1$} &
\colhead{T$_{B2}$}    &\colhead{$\sigma_2$} 
\nl
\colhead{\ }          &\colhead{(")} &
\colhead{(K)}          &\colhead{(km/s)} &
\colhead{(K)}          &\colhead{(km/s)} &
\nl
\colhead{(1)}          &\colhead{(2)} &
\colhead{(3)}          &\colhead{(4)} &
\colhead{(5)}          &\colhead{(6)} &
} 
\startdata
N5457&200& 4.7$^{+0.2}_{-0.2}$& 18.$^{+1}_{-1}$& 1.4$^{+0.6}_{-0.5}$& 
4.3$^{+2.6}_{-1.7}$ \nl
\    &600& 3.5$^{+0.4}_{-0.3}$& 13.$^{+2}_{-2}$& 1.7$^{+0.7}_{-0.6}$& 
4.5$^{+2}_{-2}$ \nl
\    &1000& 3.3$^{+0.2}_{-0.2}$& 11.$^{+0.6}_{-0.6}$&1.2$^{+0.7}_{-0.2}$& 
2.8$^{+0.9}_{-0.9}$ \nl
\    &1400& 3.7$^{+0.3}_{-0.3}$& 8.7$^{+1.0}_{-1.0}$&0.9$^{+0.9}_{-0.5}$& 
3$^{+4}_{-2}$ \nl
\enddata
\end{deluxetable}

\clearpage

\figcaption{Illustration of noise properties in the database. {\bf
    (a)} A histogram of the observed brightness in several spectral
  channels containing no discernible line emission. The solid curve is
  a Gaussian with $\sigma$~=~1.05 mJy/Beam area. The dashed curve
  corresponds to $\sigma$~=~1.07 mJy/Beam area. {\bf (b)} A histogram
  of the velocities isolated by the mask in a subsection of Fig.~2c.
  {\bf (c)} The histogram of velocities in the same region which also
  satisfy: associated pixel brightness $>~3~\sigma$. {\bf (d)} The
  histogram of velocities in the same region which also satisfy:
  associated pixel brightness $>~4~\sigma$.
\label{fig1}}

\figcaption{Illustration of data purity in the database. {\bf (a)} The
  peak brightness of \ion{H}{1} emission in NGC~5457 is shown
  (as in B95).  {\bf (b)} The same image after application of a
  spatial mask (as described in the text).  {\bf (c)} The
  line-of-sight velocities of those pixels shown in (b).  {\bf (d)} The
  line-of-sight velocities which also satisfy: associated pixel
   brightness $>~3~\sigma$.
  \label{fig2}}

\figcaption{Radial profiles and spectra of the high brightness network
  of \ion{H}{1} in NGC~247. {\bf (a)} Filled circles indicate the peak
  brightness (the 80$^{th}$ percentile) at full spatial resolution,
  open circles the kinetic temperature fit to the line profiles, open
  squares the apparent face-on mass surface density, and the solid
  line is the face-on blue light surface brightness. Open triangles
  with error bars are the isolated measurements of mean spin
  temperature made against continuum sources. Half error bars indicate
  3$\sigma$ lower limits to the mean spin temperature.  {\bf (b)}
  Spatial and velocity coherence. Filled circles indicate the peak
  brightness as in (a), the other symbols represent the median ratio
  of diminished peak brightness (see text) after smoothing to
  9~arcsec: open circles, 15~arcsec: open squares, 10~km~s$^{-1}$:
  three arm star and 20~km~s$^{-1}$: six arm star.  {\bf (c)} Filled
  circles indicate the peak brightness as in (a), open circles the
  face-on surface covering factor of the HBN, stars indicate the local
  fraction of \ion{H}{1} flux contained in the HBN.  The short-dashed
  line is the accumulated flux in the HBN normalized to the total
  flux. The long-dashed line is the normalized accumulated total flux.
  {\bf (d)} Median, co-aligned emission line profiles of brightness
  temperature within annular elliptical zones at the indicated radii.
  The solid line is the fit tabulated in Table~3.
\label{fig3}}

\figcaption{Radial profiles and spectra of NGC~2366 with panels as in Fig.~3. 
\label{fig4}}

\figcaption{Radial profiles and spectra of NGC~2403 with panels as in Fig.~3. 
\label{fig5}}

\figcaption{Radial profiles and spectra of NGC~3031 with panels as in Fig.~3. 
\label{fig6}}

\figcaption{Radial profiles and spectra of NGC~4236 with panels as in Fig.~3. 
\label{fig7}}

\figcaption{Radial profiles and spectra of NGC~5457 with panels as in Fig.~3. 
\label{fig8}}

\figcaption{Radial profiles and spectra of NGC~7793 with panels as in Fig.~3. 
\label{fig9}}

\figcaption{Median co-aligned emission line profiles of diffuse disk
  emission in NGC~5457 within annular elliptical zones at the
  indicated radii and with peak T$_B~<~7.5$~K. The expected degree of
  beam-smearing due solely to galactic kinematics (at this resolution
  of 65\arcsec) is indicated by the horizontal bar. The solid lines
  are the fits tabulated in Table~4.
\label{fig10}}

\figcaption{Radial profiles of peak brightness temperature in the
  sample galaxies. {\bf (a)} Peak brightness (the 80$^{th}$
  percentile) is plotted as a function of physical radius with
  different symbols for the different morphological types. Solid lines
  connect the points for individual galaxies. {\bf (b)} Peak
  brightness is plotted as a function of radius in units of the blue
  stellar disk scalelength.
\label{fig11}}


\clearpage
\plotone{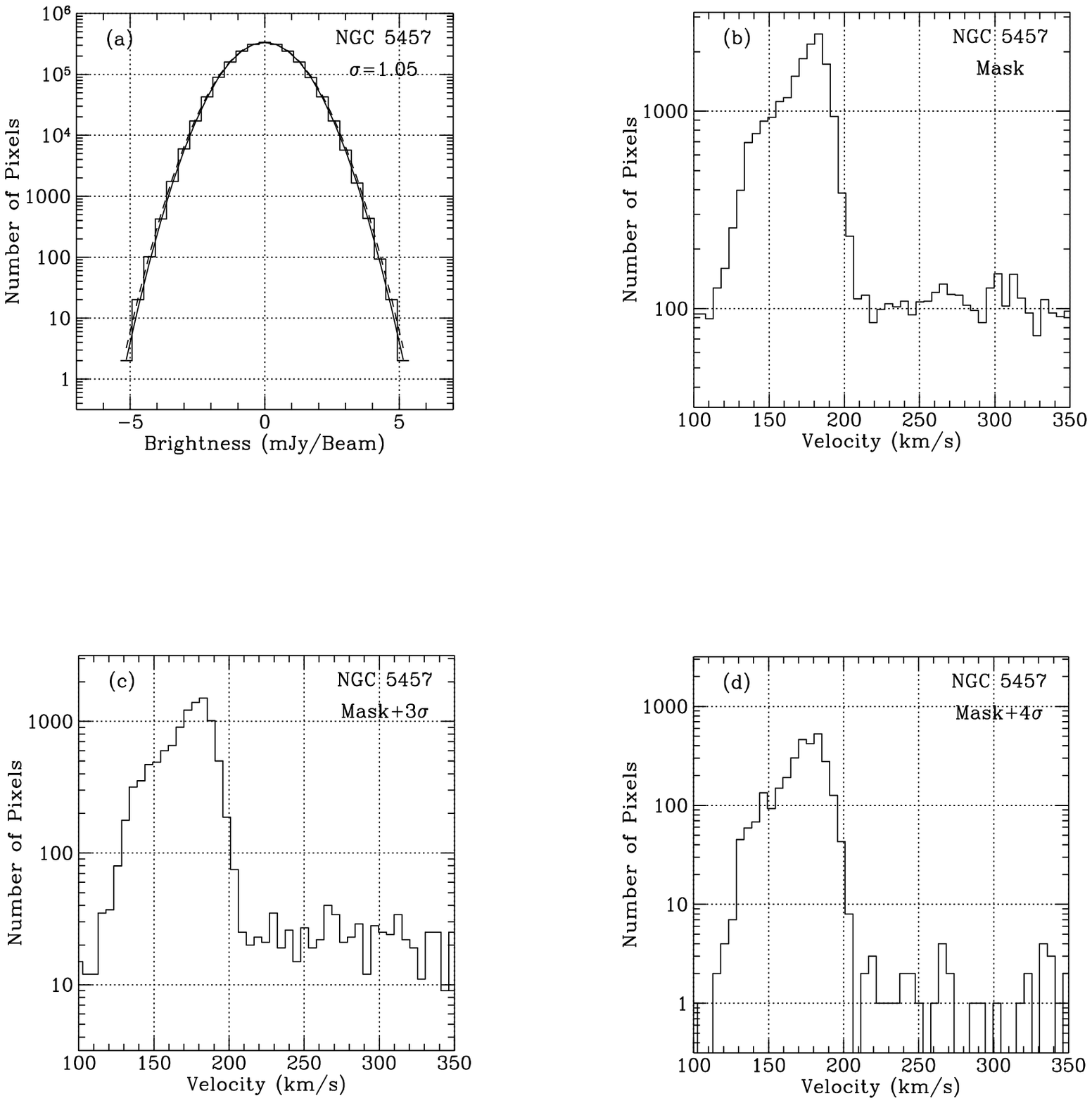}
\clearpage
\plotone{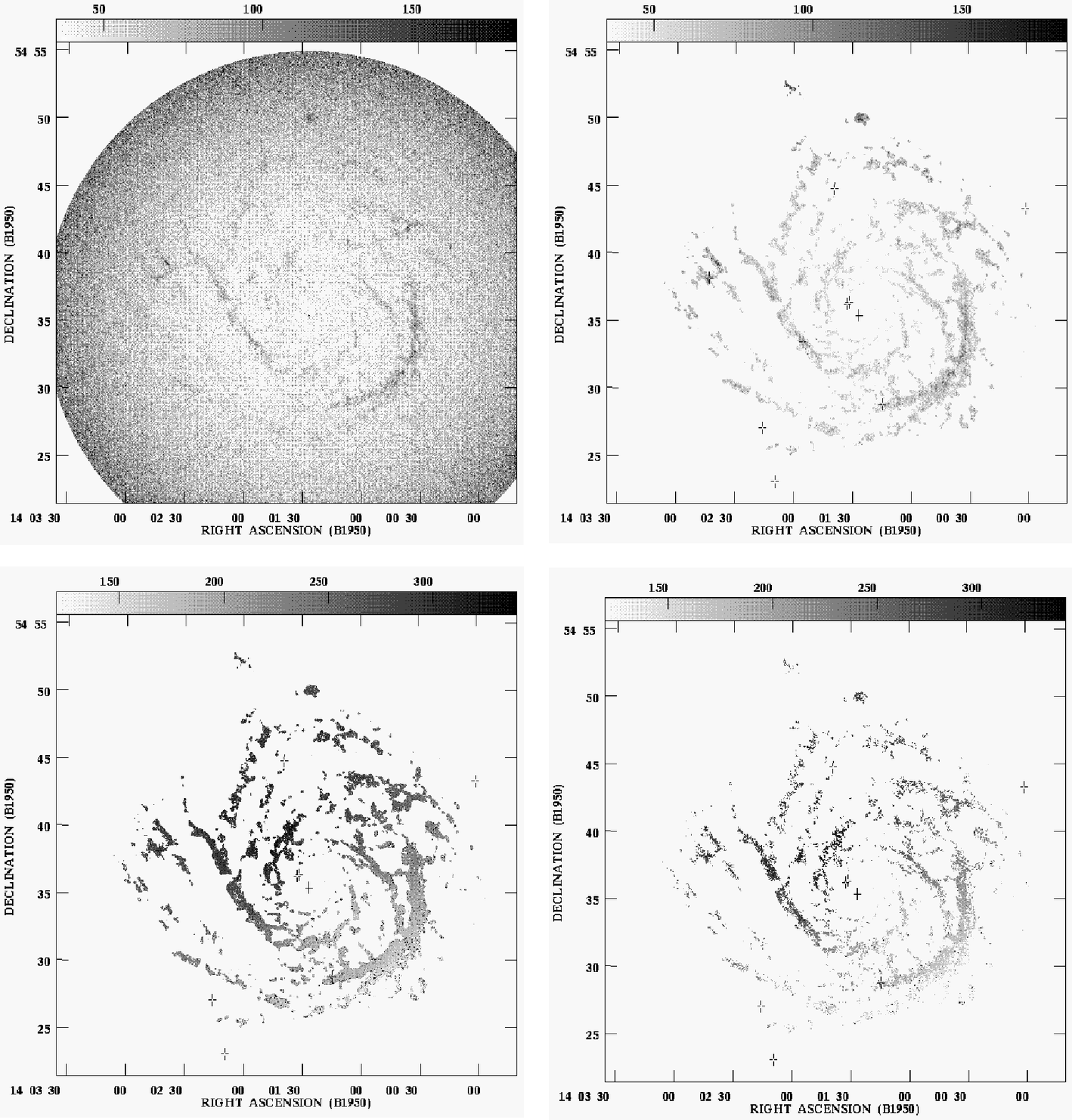}
\clearpage
\plotone{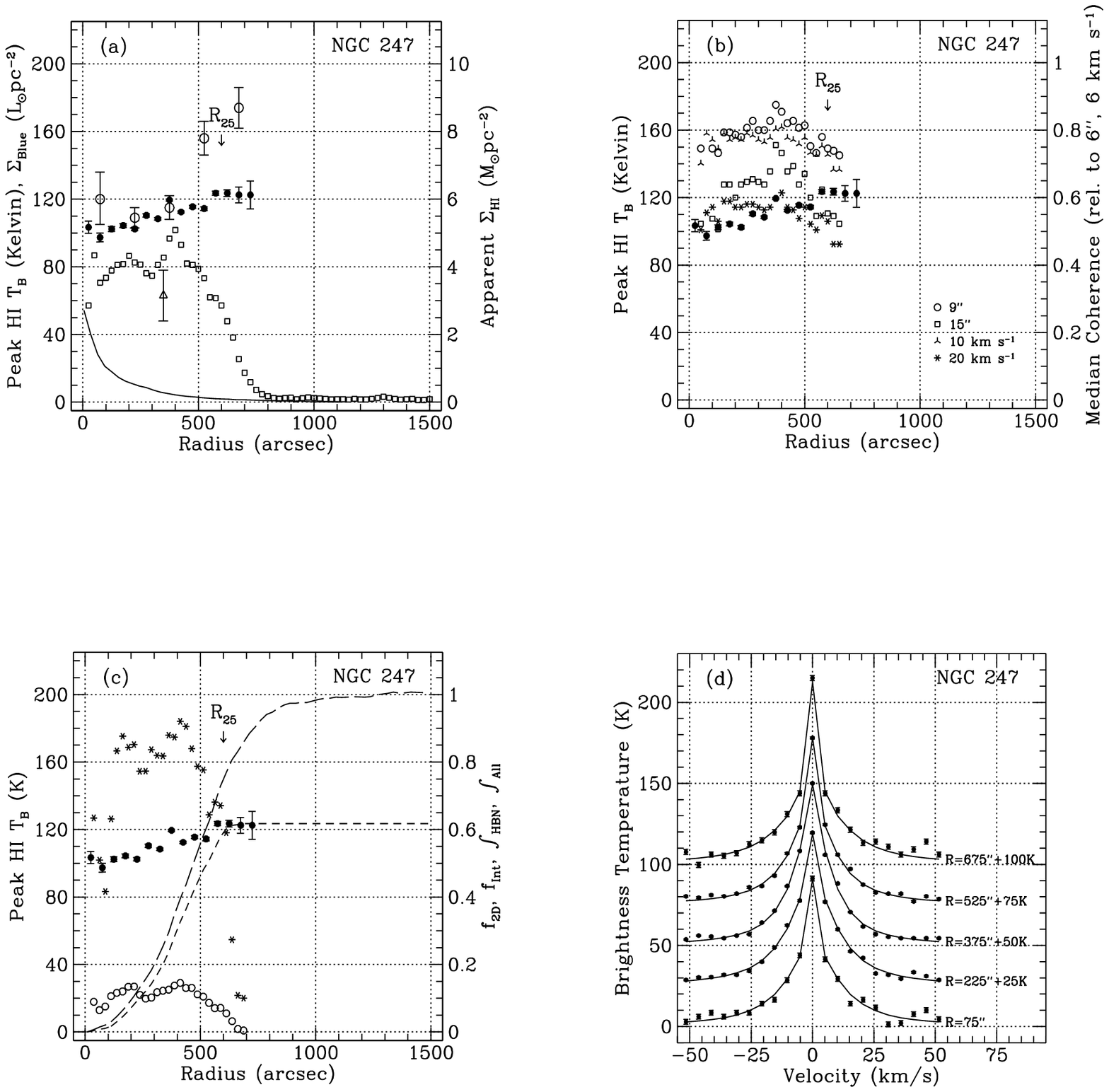}
\clearpage
\plotone{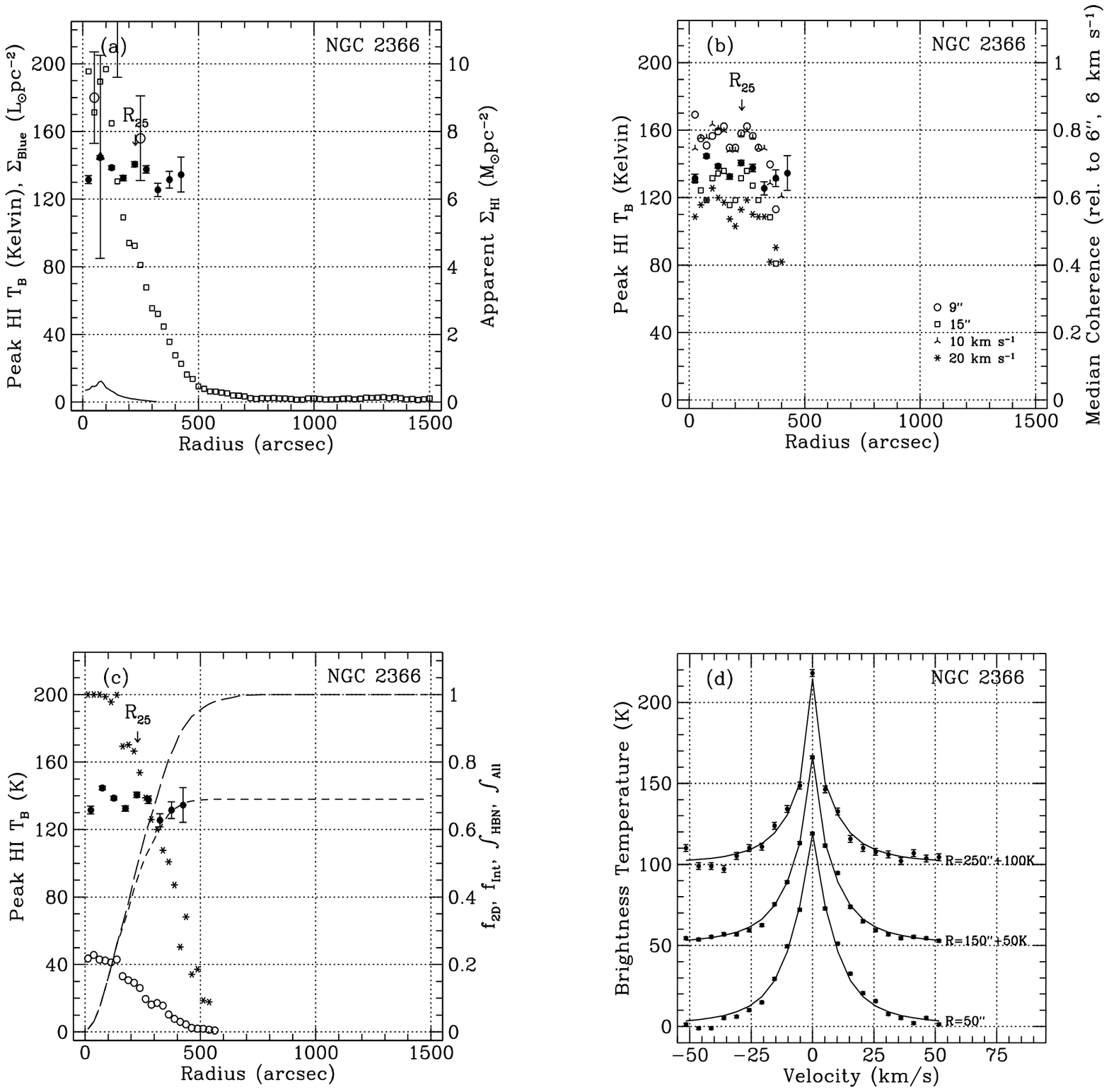}
\clearpage
\plotone{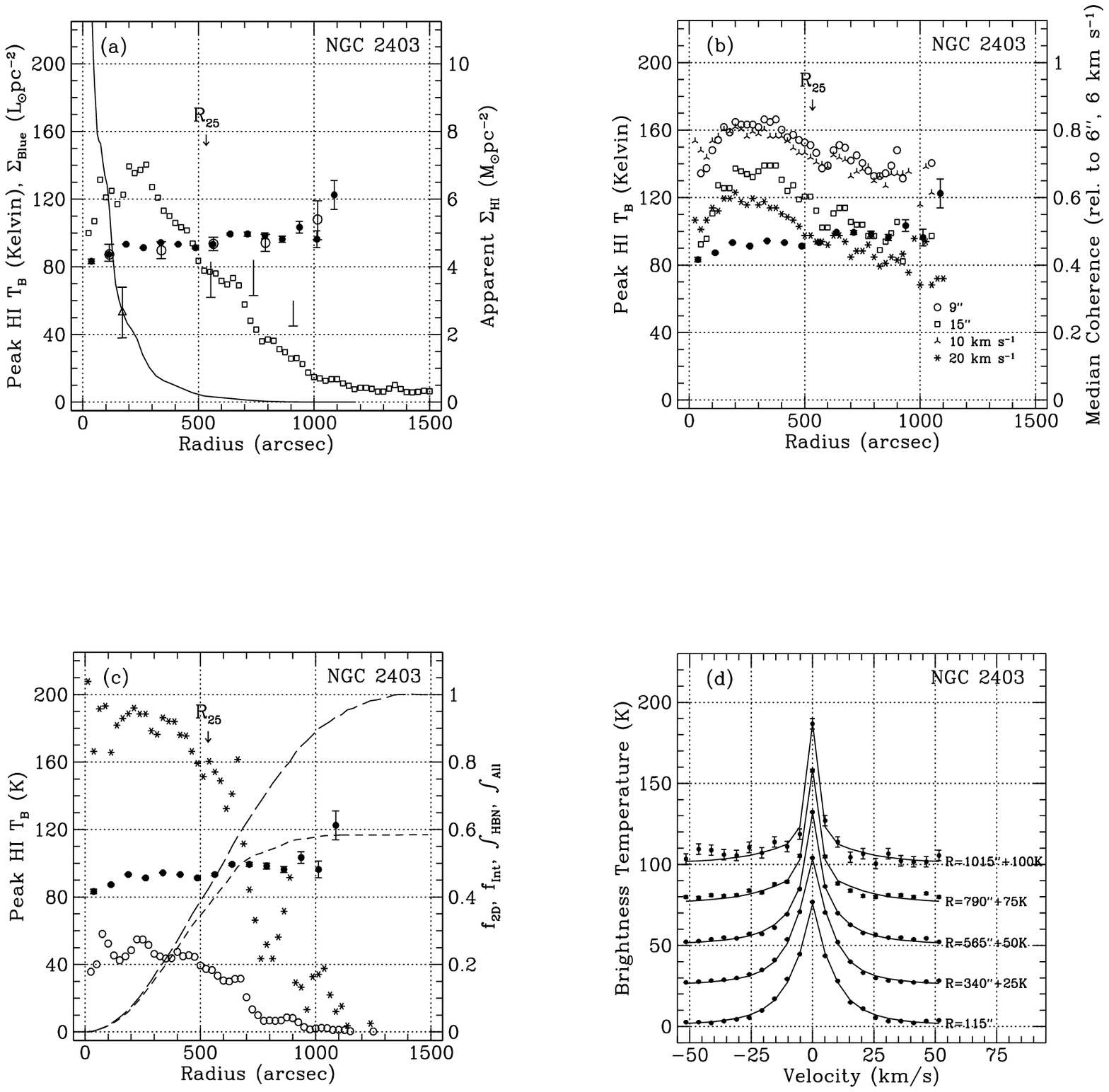}
\clearpage
\plotone{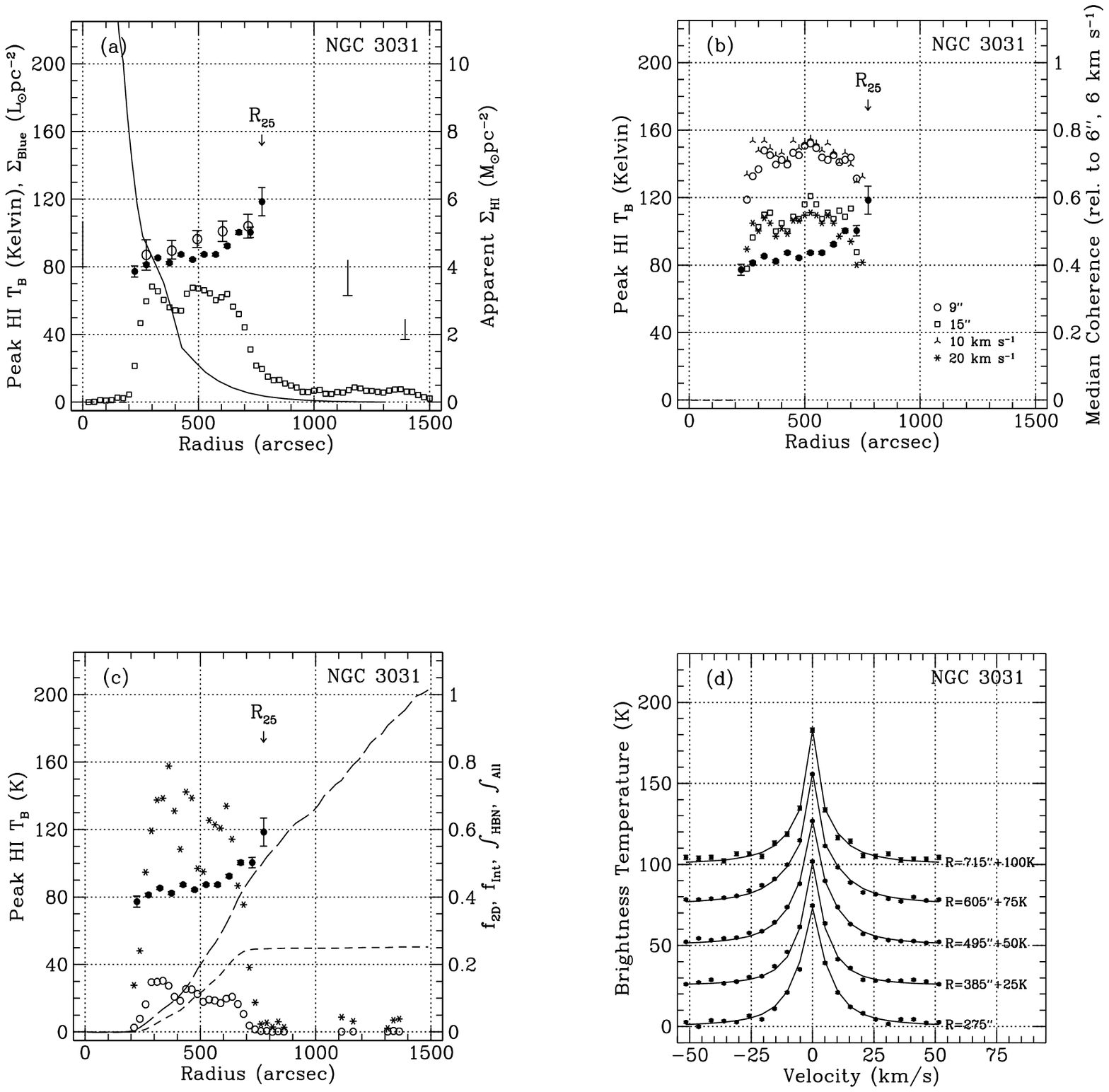}
\clearpage
\plotone{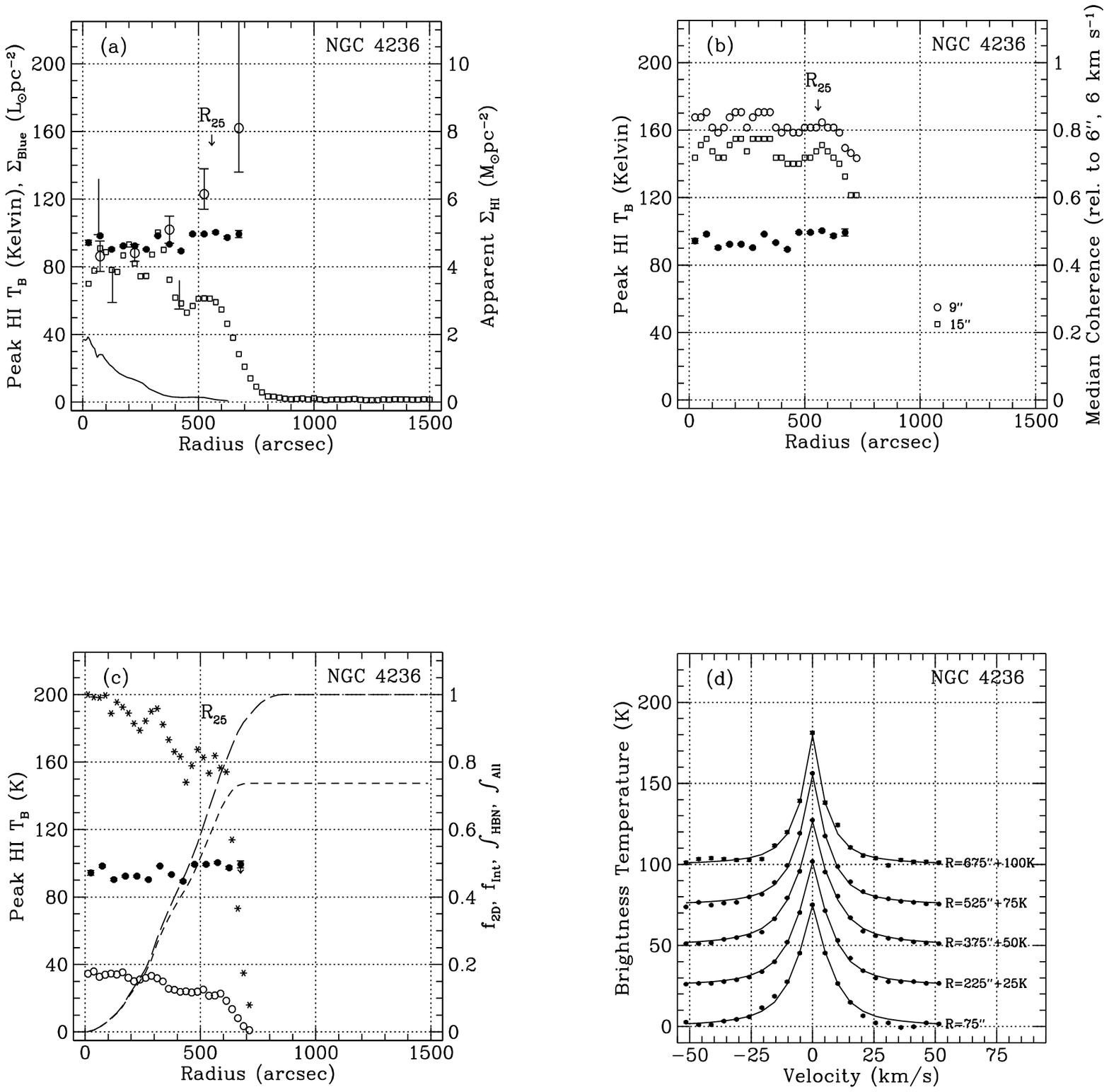}
\clearpage
\plotone{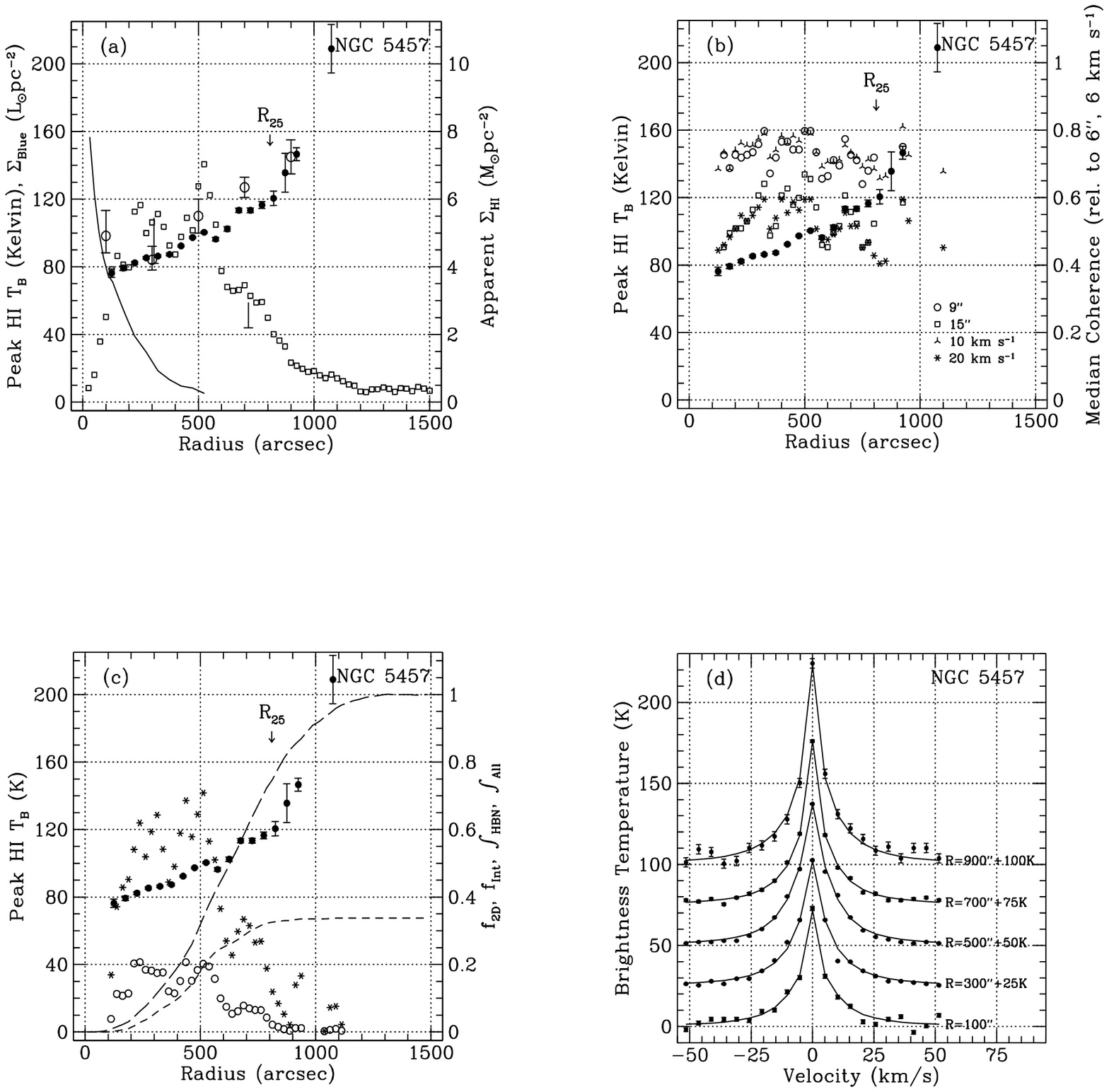}
\clearpage
\plotone{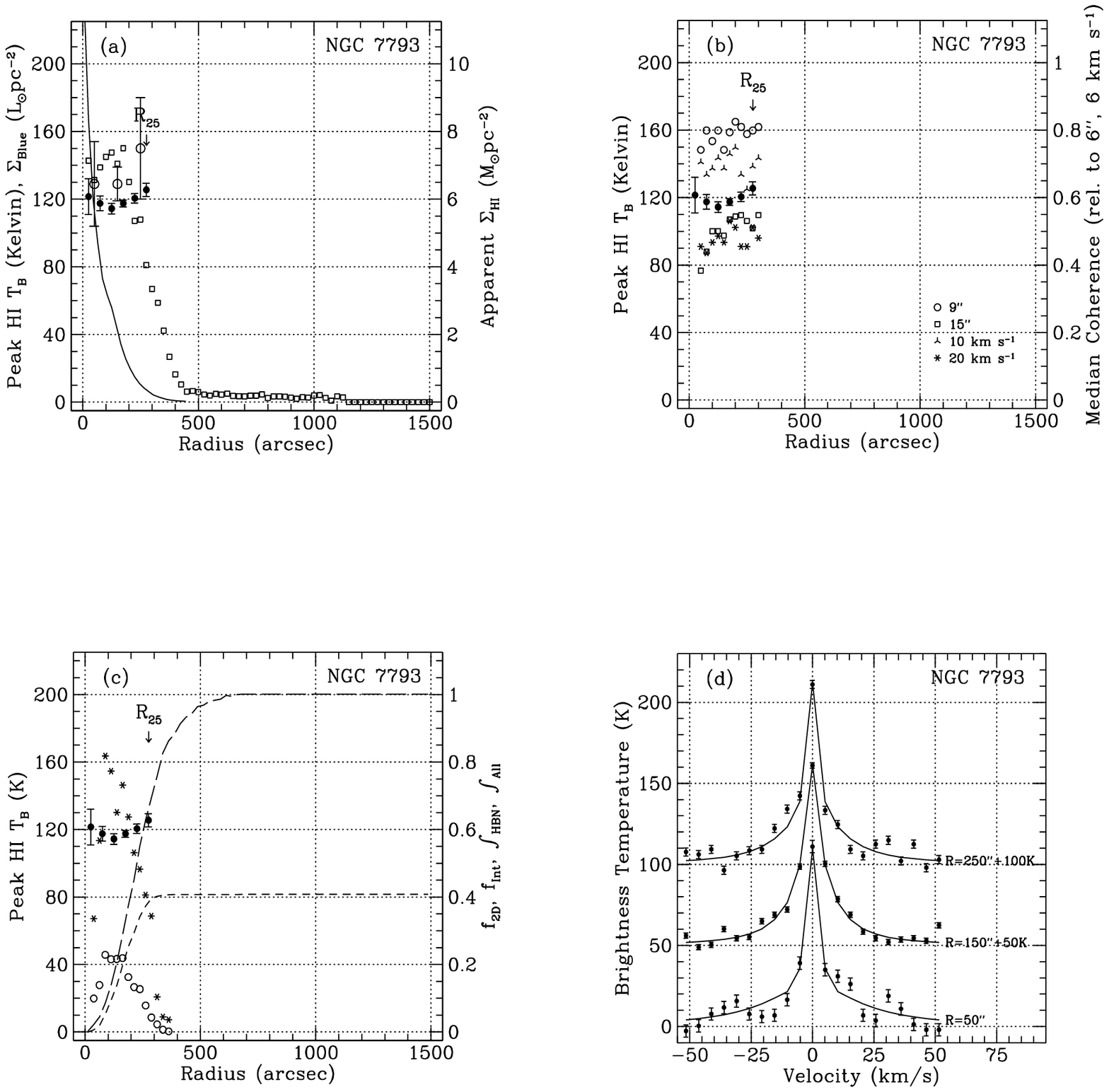}
\clearpage
\plotone{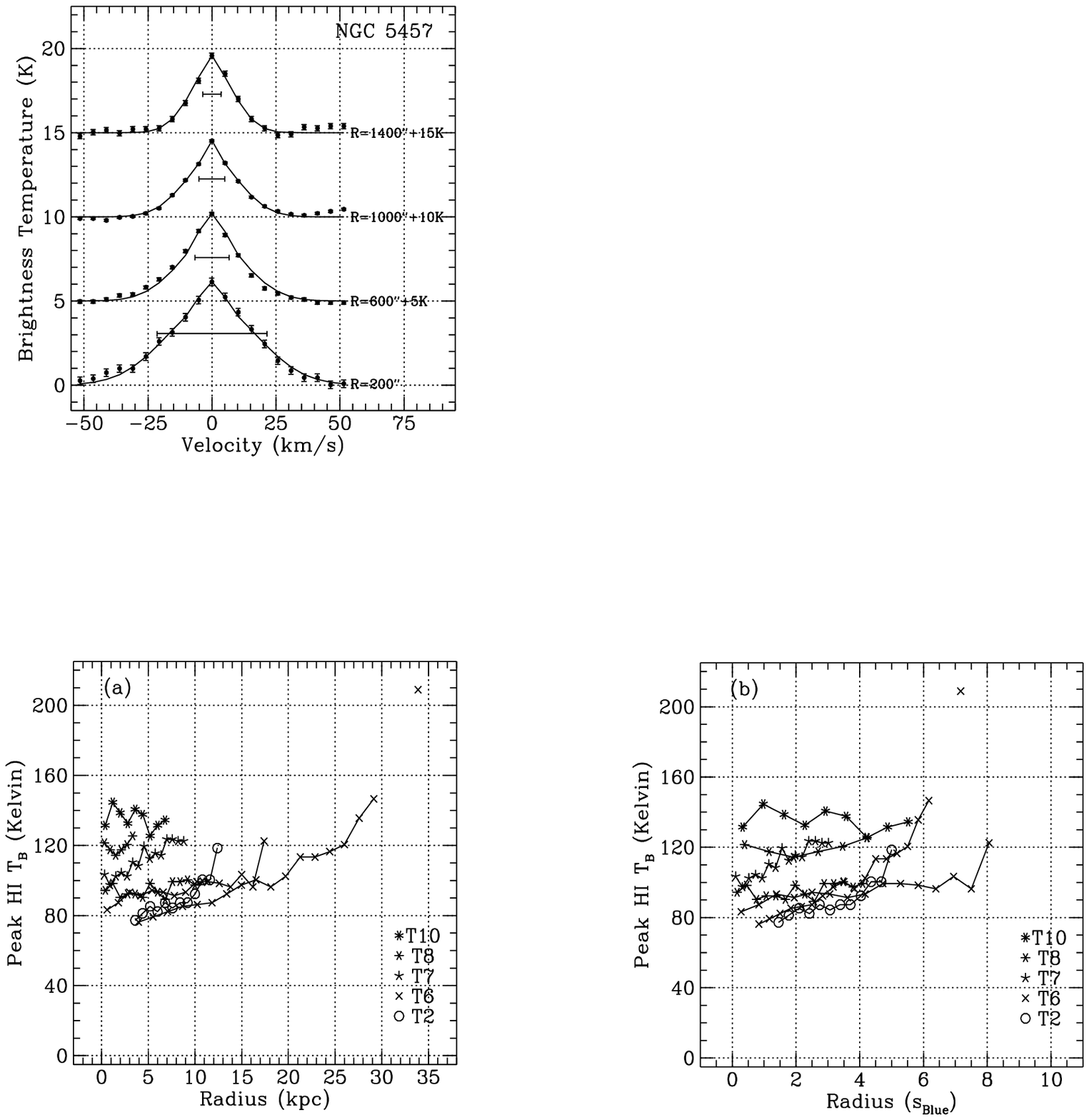}


\clearpage

\begin{thebibliography}{}

\bibitem[ ]{} 
Begeman, K.G. 1987, PhD thesis, Univ. Groningen
\bibitem[ ]{} 
Braun, R. 1995, \aaps, 114, 409 (B95)
\bibitem[ ]{} 
Braun, R., \& Walterbos, R.A.M. 1992, \apj, 386, 120
\bibitem[ ]{} 
Briggs, F.H. 1990, \apj, 352, 15
\bibitem[ ]{} 
Brinks, E., \& Shane, W.W. 1984, \aaps, 55, 179
\bibitem[ ]{} 
Clark , B.G. 1965, \apj, 142, 1398
\bibitem[ ]{} 
Deul, E.R., \& van der Hulst, J.M. 1987, \aaps, 67, 509
\bibitem[ ]{} 
de Vaucouleurs, G., de Vaucouleurs, A., Corwin, H.G. et al. 1991, Third
Reference Catalogue of Bright Galaxies (New York: Springer-Verlag) (RC3)
\bibitem[ ]{} 
Draine, B.T. 1978, \apjs, 36, 595
\bibitem[ ]{} 
Field, G.B. 1965, \apj, 142, 531
\bibitem[ ]{} 
Heiles, C. 1967, \apjs, 15, 97
\bibitem[ ]{} 
Huchtmeier, W.K., \& Richter, O.-G. 1989, A General Catalog of HI
Observations of Galaxies (New York: Springer-Verlag)
\bibitem[ ]{} 
Kulkarni, S.R., \& Fich, M. 1985, \apj, 289, 792
\bibitem[ ]{} 
Kulkarni, S.R., \& Heiles, C. 1988, in Galactic and Extragalactic Radio
Astronomy, ed. K.I. Kellerman \& G.L. Verschuur (Heidelberg:
Springer-Verlag), 95
\bibitem[ ]{} 
Radhakrishnan, V., Murray, J.D., Lockhart, P., \& Whittle, R.P.J. 1972,
\apjs, 24, 15
\bibitem[ ]{} 
Shull, J.M., \& Woods, D.T. 1985, \apj, 288, 50
\bibitem[ ]{} 
Spitzer, L., Jr. 1978, Physical Processes in the Interstellar Medium
(New York:John Wiley and Sons)
\bibitem[ ]{} 
Verschuur, G.L. 1974, \apjs, 27, 283
\bibitem[ ]{} 
Walterbos, R.A.M., \& Braun, R. 1996, in Minnesota Lectures on
Extra-Galactic \ion{H}{1}, ed. E. Skillman (ASP)
\bibitem[ ]{} 
Wolfire, M.G. , Hollenbach, D., McKee, C.F. et al. 1995, \apj, 443, 152

\end{thebibliography}
\end{document}